\documentclass[11pt]{article}
\usepackage{epsfig}
\usepackage{amsmath}
\usepackage{amssymb}
\usepackage{amsthm}
\usepackage{epstopdf}
\usepackage{graphicx}
\usepackage{enumerate}
\usepackage{authblk}
\usepackage{array}
\usepackage{cite}
\usepackage{color}
\usepackage[margin=0.9in]{geometry}
\usepackage[caption=false]{subfig}
\usepackage{float}
\normalsize \baselineskip 16pt
\newcommand {\beq} {\begin{equation}}
\newcommand {\eeq} {\end{equation}}

\newcolumntype{P}[1]{>{\centering\arraybackslash}p{#1}}
\newcolumntype{M}[1]{>{\centering\arraybackslash}m{#1}}

\newtheorem{remark}{Remark}

\begin{document}
\title{Unstable dynamics of solitary traveling waves in a lattice with long-range interactions}
\author[1]{Henry Duran}
\author[2]{Haitao Xu}
\author[3]{Panayotis\ G.\ Kevrekidis}
\author[1]{Anna Vainchtein}
\affil[1]{\small Department of Mathematics, University of Pittsburgh, Pittsburgh, Pennsylvania 15260, USA}
\affil[2]{\small Center for Mathematical Science, Huazhong University of Science and Technology, Wuhan, Hubei 430074, People's Republic of China}
\affil[3]{\small Department of Mathematics and Statistics, University of
Massachusetts, Amherst, MA 01003-9305, USA}

\maketitle

\begin{abstract}
  In this work we revisit the existence, stability and dynamics of
  unstable traveling solitary waves in the context of lattice dynamical
  systems. We consider a nonlinear lattice of an $\alpha$-Fermi-Pasta-Ulam type
  with the additional feature of all-to-all harmonic long-range interactions whose strength decays exponentially with distance.
  The competition between the nonlinear nearest-neighbor
  terms and the longer-range linear ones yields two parameter regimes where the dependence of the energy $H$
  of the traveling waves on their velocity $c$ is non-monotonic and multivalued, respectively.
  We examine both cases, and identify the exact (up to a prescribed numerical tolerance) traveling waves.
  To investigate the stability of the obtained solutions, we compute their Floquet multipliers, thinking of the traveling
  wave problem as a periodic one modulo shifts. We show that in the general case when the relationship between $H$ and $c$
  is not single-valued, the sufficient but not necessary criterion for stability change is $H'(s)=0$, where $s$ is the parameter
  along the energy-velocity curve. Perturbing the unstable solutions along the corresponding eigenvectors,
  we identify two different scenarios of the dynamics of their transition to stable branches.
  In the first one, the perturbed wave slows down after expelling a dispersive wave. The second scenario involves an increase in the velocity
  of the perturbed wave accompanied by the formation of a slower small-amplitude traveling solitary wave.
\end{abstract}

\noindent {\bf Keywords:} lattice dynamics, long-range interactions, traveling solitary wave, Floquet spectrum, instability, energy-based criterion

\section{Introduction}
Since the groundbreaking work~\cite{FPU55,ZK65} on nonlinear Fermi-Pasta-Ulam (FPU) lattices, among the principal objects of investigation have been the solitary traveling waves (STWs) that emerge therein and their connection to soliton solutions of the Korteweg-de Vries (KdV) equation. Consequently, many studies have been devoted to understanding the properties of these waves in discrete systems, including experimental investigations in electrical networks \cite{HirotaSuzuki73,KMR88}, granular materials \cite{CFF97,Nesterenko01}, and more recently in mechanical metamaterials \cite{Deng19b,Yasuda19} and lipid monolayers \cite{Shrivastava15}. Significant theoretical developments include the discovery of the integrable Toda lattice and the study of its  STWs~\cite{Toda81}, existence proofs for non-integrable systems \cite{FW94,SW97,PR11,Stefanov12} and rigorous investigations of the low-energy \cite{FP99,FP02,FP04a,FP04b,Iooss00,McMillan02,Hoffman13} and high-energy \cite{FM02,HerrmannMatthies15,HerrmannMatthies19} limits.

Despite all this progress, stability of lattice STWs remains an issue that is far from being fully understood, with rigorous results only known for some special cases such as the integrable Toda lattice \cite{Mizumachi08,Benes12}, near-integrable sonic limit \cite{FP99,FP02,FP04a,FP04b} and the hard-sphere high-energy limit \cite{HerrmannMatthies19}. A sufficient condition for change in the spectral stability of a STW was established in \cite{FP04a} for the FPU problem. In the recent work \cite{Cuevas17,Xu18} this result was extended to a general class of Hamiltonian systems and
connected to stability criteria in the realm of discrete breathers~\cite{Cuevas16}. This energy-based criterion involves the monotonicity of the Hamiltonian $H$ as a function of the wave's velocity $c$. The corresponding criterion for breathers, time-periodic localized solutions, concerns the monotonicity of $H$ with respect to the frequency $\omega$ of the breather. The intimate connection between the criteria stems from the fact that traveling waves are periodic
modulo lattice shifts, resulting in the direct proportionality of $\omega$ and $c$.
The relevant stability criterion states that as $c$ is varied, passing through a critical point of $H(c)$ is sufficient (but not necessary) for a change in stability. As shown in \cite{Cuevas17,Xu18}, a pair of eigenvalues associated with the STW collides at zero at the critical velocity value and reemerges on the real axis when the wave becomes unstable.

The combination of this criterion and the fact that STWs in the FPU problem are stable near the sonic limit, where $H'(c)>0$ \cite{FP04b} suggests that waves become unstable when $H'(c)<0$. Interestingly, in most known cases $H(c)$ is a monotonically increasing function and numerical (or, in the case of Toda lattice, analytical \cite{Mizumachi08,Benes12}) results indicate stability of all STWs. Examples of lattices with nonmonotone $H(c)$ include ones with piecewise quadratic interaction potentials \cite{TV14,TV19,KatzGivli18,KatzGivli19} and their smooth approximations \cite{Xu18}. Another remarkable example was revealed in a series of papers \cite{NGFM94,GFNM95,GFNM97,MGM00} that investigated a system with nonlinear nearest-neighbor interactions and harmonic Kac-Baker longer-range ones. In these works, the authors showed that depending on the parameters of the long-range interactions and due to an interplay of two different length scales, $H(c)$ can be monotonically increasing, nonmonotone or fold on itself ($Z$-shaped), becoming multivalued in a certain velocity interval, where three STWs with the same velocity coexist \cite{MGM00}. Numerical simulations in \cite{MGM00} suggest stability of the low-energy and high-energy solutions where $H'(c)>0$ and instability of the intermediate ones. For the nonmonotone single-valued case, this conjecture is supported by the stability analysis of the associated quasicontinuum model in \cite{GFNM97} and linear stability analysis of the discrete system in \cite{Cuevas17,Xu18} which reveals the above mentioned instability picture
associated with real eigenvalues at the spectral analysis level.

In this work we revisit this problem and extend the analysis in \cite{Cuevas17,Xu18} to the case when $H(c)$ is no longer single-valued. We show that the change of stability is now associated with the change of sign of $H'(s)$, where $s$ is the parameter that $c$ and $H$ depend on. Representing STWs as periodic-modulo-shift orbits \cite{FP04a,Cuevas17,Xu18}, we perform Floquet analysis in the parameter regime where $H(c)$ is $Z$-shaped and show that instability in this case is associated with $H'(s)<0$. In the case of nonmonotone $H(c)$ this reduces to $H'(c)<0$.

A related central scope of this work is to
investigate in detail the dynamical
consequences of instability in both of these regimes.
We do this by perturbing the unstable waves along the eigenmode corresponding to a real Floquet multiplier associated with the instability and tracking the velocity and energy of the evolving wave. Our results show that depending on the sign of perturbation, there are two generic scenarios. In the first case, the wave slows down after expelling a dispersive shock wave. In the second scenario, the wave's velocity increases following the formation and expulsion of a small-amplitude STW. In both cases, the waves stabilize when their velocity reaches a value along the energy-velocity curve where $H'(c)>0$.

The remainder of the paper is organized as follows. In Sec.~\ref{sec:prior} we formulate the problem and review prior results. In Sec.~\ref{sec:numer_methods} we describe the numerical methods we used. Results for the single-valued nonmonotone $H(c)$ are presented in Sec.~\ref{sec:Nregion}, while Sec.~\ref{sec:Zregion} is devoted to the multivalued case. Concluding remarks can be found in Sec.~\ref{sec:conclusions}. A more technical stability analysis for multivalued $H(c)$ is presented in the Appendix.

\section{Problem formulation and prior results}
\label{sec:prior}
We consider Hamiltonian dynamics of a one-dimensional lattice with nonlinear nearest-neighbor interactions and all-to-all harmonic longer-range interactions, with moduli that decay exponentially with distance. The Hamiltonian of this system is given by
\beq
H = \sum_{n=-\infty}^{\infty} \left\{\frac{1}{2}\dot{u}_n^2+V(u_{n+1}-u_n)
+\frac{1}{4}\sum_{m=-\infty}^{\infty}\Lambda(m)(u_n-u_{n+m})^2 \right\},
\label{eq:Ham}
\eeq
where $u_n(t)$ denotes the displacement of $n$th particle at time $t$, $\dot{u}_n=u_n'(t)$, and $V(w)=w^2/2-w^3/3$ is the potential governing the nearest-neighbor interactions. The last term represents Kac-Baker  interactions that have moduli $\Lambda(m)=J(e^{\alpha}-1) e^{-\alpha |m|}(1-\delta_{m,0})$. Here $J$ measures the intensity of the longer-range interactions, and $\alpha$ determines their inverse radius. In terms of strain (relative displacement) variable $w_n=u_{n+1}-u_n$, equations of motion are
\beq
\ddot{w}_n+2V'(w_{n})-V'(w_{n+1})-V'(w_{n-1})+\sum_{m=1}^{\infty}\Lambda(m)(2w_n-w_{n+m}-w_{n-m})=0.
\label{eq:EoM}
\eeq
The energy $H$ and the total momentum
\beq
P=\sum_{n=-\infty}^{\infty} \dot{u}_n
\label{eq:P}
\eeq
of the system are conserved in time.

Previous work \cite{NGFM94,GFNM95,GFNM97,MGM00,Cuevas17,Xu18} on this model has focused on \emph{solitary traveling wave} (STW) solutions of \eqref{eq:EoM}, which have the form
\beq
w_n(t)=\phi(\xi), \quad \xi=n-ct,
\label{eq:STW_ansatz}
\eeq
where $c$ is the wave's velocity, and vanish at infinity. These solutions satisfy the advance-delay differential equation
\beq
c^2\phi''(\xi)+2V'(\phi(\xi))-V'(\phi(\xi+1))-V'(\phi(\xi-1))
+\sum_{m=1}^\infty \Lambda(m)(2\phi(\xi)-\phi(\xi+m)-\phi(\xi-m))=0.
\label{eq:STW_eqn}
\eeq
Numerical computations in \cite{Cuevas17,GFNM97,MGM00,Xu18} suggest the existence of even ($\phi(-\xi)=\phi(\xi)$), compressive ($\phi(\xi)<0$) solutions of this type with $c>c_\text{s}$, where
\beq
c_\text{s}=\sqrt{1+J\dfrac{1+e^{-\alpha}}{(1-e^{-\alpha})^2}}
\label{eq:sound}
\eeq
is the sound speed \cite{MGM00}. Due to the translational invariance of \eqref{eq:STW_eqn}, these waves can be shifted arbitrarily along the $\xi$ axis. Note also that the traveling wave solutions \eqref{eq:STW_ansatz} are periodic modulo one lattice shift, $w_{n+1}(t+T)=w_n(t)$,
with period $T=1/c$, and thus can be viewed as fixed points of the map
\begin{equation}\label{eq:map}
\left[\begin{array}{c}
  \{w_{n+1}(T)\} \\ \{\dot w_{n+1}(T)\} \\  \end{array}\right]
  \rightarrow
  \left[\begin{array}{c}
  \{w_{n}(0)\} \\ \{\dot w_{n}(0)\} \\  \end{array}\right].
\end{equation}

In \cite{GFNM97,MGM00} the lattice equations \eqref{eq:EoM} are approximated by a quasicontinuum model, in which  $n$ is replaced by $x$, $w_{n}(t)$ by $w(x,t)$ and $w_{n \pm m}(t)$ by $e^{\pm m \partial_x}w(x,t)$. Using Taylor expansion of the shift operators $e^{\pm m \partial_x}$, one obtains
\beq
[{\partial_{t}}^2-JQ(\alpha,\partial_x)]w(x,t)-4\text{sinh}^2\left(\frac{\partial_x}{2}\right)V'(w)=0.
\label{eq:quasi}
\eeq
where
\beq
Q(\alpha,\partial_x)=(e^{\alpha}+1)\frac{4\text{sinh}^2(\partial_{x}/2)}{\kappa^2-4\text{sinh}^2(\partial_x/2)}
\label{eq:Oper}
\eeq
is a linear pseudodifferential operator.

Detailed analysis of the quasicontinuum approximation \eqref{eq:quasi} in \cite{GFNM97} (see also \cite{GFNM95}) has shown that the interplay of short-range and long-range interactions in the problem gives rise to two competing velocity-dependent length scales. In a certain parameter regime, this scale competition leads to the existence of two branches of STWs, associated with low and high velocities, respectively, and the emergence of crest-like waves when the velocity reaches a critical value.

\begin{figure}[!htb]
\centerline{\psfig{figure=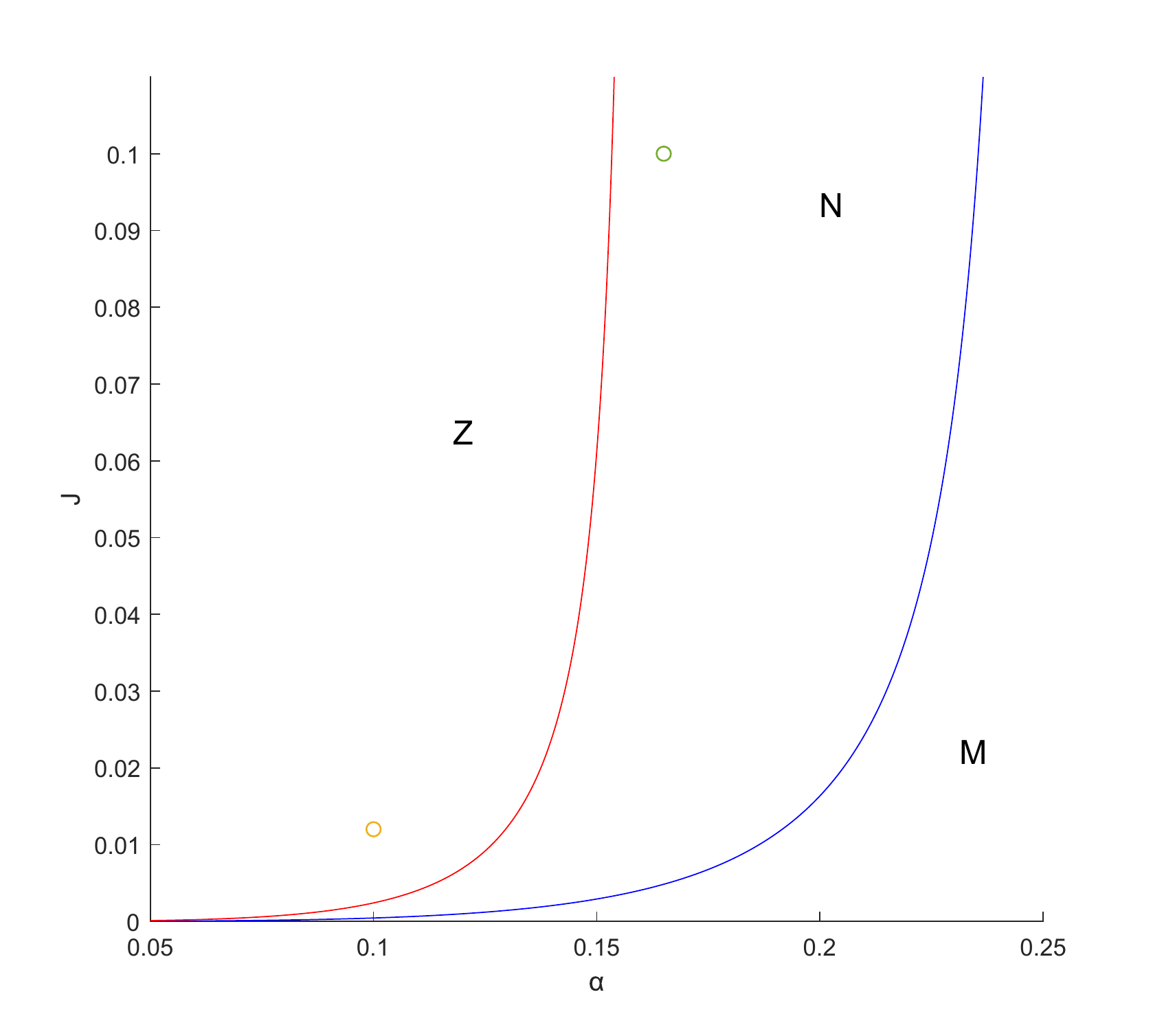,width=0.4\textwidth}}
\caption{\footnotesize The $M$, $Z$ and $N$-regions in the $(\alpha,J)$ plane together with the boundary curves $J_1(\alpha)$ (right) and $J_2(\alpha)$ (left) defined in \eqref{eq:kacbakercond}. Circles mark the parameter values for the examples discussed in Sec.~\ref{sec:Nregion} and Sec.~\ref{sec:Zregion}.}
\label{fig:Region}
\end{figure}

Numerical computations in \cite{MGM00} of solutions of \eqref{eq:STW_eqn} for the discrete problem further showed that the $(\alpha,J)$ plane can be subdivided into three regions, separated by the curves $J_1(\alpha)$ and $J_2(\alpha)$, where
\beq
\label{eq:kacbakercond}
    J_1 \approx \begin{cases}0.23\dfrac{\alpha^4}{\alpha_1^2-\alpha^2}, & \alpha<\alpha_1\\
    \infty, & \alpha \geq \alpha_1,
    \end{cases}
    \qquad
    J_2 \approx \begin{cases}\dfrac{3\alpha^4}{8(\alpha_2^2-\alpha^2)}, & \alpha<\alpha_2\\
    \infty, & \alpha \geq \alpha_2
    \end{cases}
\eeq
and $\alpha_1=0.25$, $\alpha_2=0.16$. These three regions are shown in Fig.~\ref{fig:Region}. They consist of
the $M$-region, where the energy $H$ of the STW monotonically increases with its velocity $c$, the $N$-region, where the dependence is \emph{nonmonotone}, with $H(c)$ initially increasing, then decreasing for a certain velocity interval and then increasing again, and lastly the $Z$-region, where the function $H(c)$ becomes \emph{multivalued}  for some velocities (``$Z$-shaped''). The three different regimes were also captured in \cite{NGFM94} using a collective-coordinate approach.

It has been conjectured in \cite{MGM00} that in the $N$ and $Z$-regions the low-velocity and high-velocity solutions where $H'(c)>0$ are stable, while waves along the intermediate branch are unstable. These assertions are supported by the stability analysis in \cite{GFNM97} for the quasicontinuum model \eqref{eq:quasi}, where the stability threshold is linked to the change of monotonicity of the canonical momentum as the function of the velocity $c$ of the wave, which appears to coincide with the corresponding change in the monotonicity of $H(c)$. In \cite{FP04b} an analogous energy-based stability criterion, associating the change in stability with the change of sign of $H'(c)$, was proved for the FPU problem without long-range interactions, and in \cite{Cuevas17,Xu18} this result was extended to a general class of discrete systems with Hamiltonian $H$ being a single-valued function of $c$. Moreover, explicit leading-order expressions for the pertinent pair of eigenvalues that meet at
the origin at the stability threshold and emerge on the real axis at velocity values corresponding to the unstable waves were obtained in \cite{Cuevas17,Xu18}. For the problem at hand, this general result was illustrated in \cite{Cuevas17,Xu18} by considering STWs in the $N$-region and investigating linear stability in two different ways: the spectral analysis of the linear operator associated with the traveling wave equation \eqref{eq:STW_eqn} and the Floquet analysis of the linearization of the map \eqref{eq:map}. Both approaches corroborated the conjecture in \cite{MGM00} for the $N$-region. In particular, the waves corresponding to $H'(c)<0$ are unstable.

In what follows we investigate in detail the consequences of this instability by perturbing the unstable STWs along the corresponding Floquet eigenvectors. To extend these results to the $Z$-region, where $H$ is a multivalued function of $c$, we generalize the energy-based stability result in \cite{Cuevas17,Xu18} and show that in this case the instability threshold is associated with $H'(s)$ crossing zero, where $s$ is a parameter that both $H$ and $c$ depend on (see the Appendix for the proof). We verify this result and the conjecture in \cite{MGM00} by conducting the Floquet analysis in the $Z$-region and investigate the corresponding unstable dynamics of STWs associated with $H'(s)<0$.

\section{Numerical methods}
\label{sec:numer_methods}
To compute the STWs in the $N$-region for given $J$ and $\alpha$, we employ  the collocation method and continuation approach described in \cite{Cuevas17,Xu18} to generate a one-parameter family of STWs (parametrized by the velocity $c$) by numerically solving the traveling wave equation \eqref{eq:STW_eqn} for STW solutions starting at an initial velocity just above the sound speed \eqref{eq:sound} and using the near-sonic solution of the quasicontinuum equation \eqref{eq:quasi} as an initial guess.  These waves are computed on the finite interval $(-L/2,L/2]$ with mesh size $\Delta \xi$ at the collocation points $\xi_j=j \Delta \xi$, $j=-N/2+1,\dots N/2$, where $N$ is even and $L=N\Delta \xi$. The fast Fourier transform is used to approximate the second-order derivative term in \eqref{eq:STW_eqn}, while the advance and delay terms $\phi(\xi \pm m)$ are evaluated at the corresponding collocation points that are well defined on the chosen mesh. Following \cite{Xu18}, we used $L=800$ and $\Delta\xi=0.1$ for a typical computation. The resulting nonlinear system is solved numerically for each velocity value using the Newton iteration method.

To compute the STWs in the $Z$-region, where the energy $H$ is multivalued for some velocities, we combine the numerical procedure described above with the pseudo-arclength continuation method \cite{Keller86} to traverse the turning points in the energy-velocity curve. In this case the traveling wave solution and its velocity $c$ depend on the arclength-like parameter $s$. In this parameter range, we used $L=1200$ and $\Delta \xi=0.1$.

To investigate linear stability of the computed waves, we use  Floquet analysis. To this end, we trace the time evolution of a small perturbation $\epsilon y_n(t)$ of the periodic-modulo-shift traveling wave solution $\hat{w}_n(t)=\phi(n-ct)$, where we recall \eqref{eq:STW_ansatz}. This perturbation is introduced in (\ref{eq:EoM}) via $w_n(t)=\hat{w}_n(t)+\epsilon y_n(t)$. The resulting $O(\epsilon)$ equation reads
\begin{equation}\label{eq:perturb}
    \ddot y_n+2V''(\hat{w}_n)y_n-V''(\hat{w}_{n+1})y_{n+1}-V''(\hat{w}_{n-1})y_{n-1}+\sum_{m=1}^\infty\Lambda(m)(2y_n-y_{n+m}-y_{n-m})=0.
\end{equation}
Then, in the framework of Floquet analysis, the stability properties of periodic orbits are resolved by diagonalizing the monodromy matrix $\mathcal{F}$ (representation of the Floquet operator in finite systems), which is defined as
\begin{equation}\label{eq:Floquet}
\left[\begin{array}{c}
  \{y_{n+1}(T)\} \\ \{\dot y_{n+1}(T)\} \\  \end{array}\right]
  =\mathcal{F}
  \left[\begin{array}{c}
  \{y_{n}(0)\} \\ \{\dot y_{n}(0)\} \\  \end{array}\right],
\end{equation}
where we recall that $T=1/c$. We remark that the Floquet operator can be equivalently constructed in terms of the perturbations of strain and momenta variables, which is consistent with the formulation considered in the Appendix. For the symplectic Hamiltonian systems we consider in this work, the linear stability of the solutions requires that the monodromy eigenvalues $\mu$ (also called Floquet multipliers) lie on the unit circle. The presence of a multiplier satisfying $|\mu|>1$ indicates an instability.

The Floquet multipliers $\mu$ are related to the eigenvalues $\lambda$ of the operator associated with the linearized problem via $\mu=e^{\lambda/c}$, so that
the eigenvalue satisfying ${\rm Re}(\lambda)>0$ corresponds to an instability. As we will show, the instability takes place when $H'(s)<0$, where $s$ is the parameter along the energy-velocity curve. In the case when $H(c)$ is single-valued, as in the $N$-region, this simplifies to $H'(c)<0$ \cite{Cuevas17,Xu18}. To find the Floquet multipliers, we construct the monodromy matrix using the numerical solution of \eqref{eq:perturb} with periodic boundary conditions.

To investigate the unstable dynamics, we perturb the wave along the unstable eigenmode, setting the initial conditions
$w_n(0)=\phi(n-n_0)+\epsilon y_{n-n_0}$ and $\dot{w}_n(0)=-c\phi'(n-n_0)+\epsilon z_{n-n_0}$ for $|n-n_0| \leq L/2$, and
$w_n(0)=\dot{w}_n(0)=0$ for $1 \leq n < n_0-L/2$ and $n_0+L/2 < n  \leq N$, with the typical eigenmode profiles for $y_n$ and $z_n$ being depicted in Fig.~\ref{fig:Eigenmodes} and $\epsilon$ measuring the strength of the applied perturbation. Here we recall that $L$ is the length of the interval on which the traveling wave $\phi(\xi)$ is numerically computed, with (even) $L$ chosen large enough for the wave to decay sufficiently at the end; typically, we set $L=800$. The computed wave is shifted by $n_0$ and padded by zeros so that the initial condition defined at $n=1,\dots,N$ has compact support. Here $n_0$ and $N$ are chosen so that the ensuing waveforms can propagate for sufficiently long time without boundary effects. Typically, we set $n_0=701$ and $N=4001$. The equations of motion \eqref{eq:EoM} are then solved numerically with this initial condition and periodic boundary conditions to investigate the fate of the unstable solution.

Of particular interest is the velocity of the ensuing waveform as a function of time. Recall that an STW solution (shifted by $n_0$) has the form $w_n(t)=\phi(n-n_0-ct)$, so that if $t_1$ and $t_2$ are such that $w_{n_1}(t_1)=w_{n_2}(t_2)=\phi(0)$, we have $c=(n_2-n_1)/(t_2-t_1)$. Here $t_1$ and $t_2$ correspond to the times when the minimum value of the STW reaches the corresponding particles $n_1$ and $n_2$. In the case of unstable dynamics, the wave is no longer steady, as its velocity and form change with time, but locally these changes are small.
With this in mind, we determine the times $t_i^*$ at which the minimum of the
waveform reaches the particle with $n_i=n_0+i\Delta n$, $i=1,\dots,K$, and approximate $c(t_i^*)$ by
\beq
c_i=\frac{\Delta n}{t_{i+1}^*-t_i^*}.
\eeq
Here $n_K$ is the particle number reached by the wave near the end of the simulation. To compute $t_i^*$ more precisely, we use cubic spline interpolation of the numerical data. Experimentally, we found that setting $\Delta n=5$ was optimal, since this value provided some averaging and yielded final velocities that were the same up to $O(10^{-5})$ as the computations with $\Delta n=3$  and $\Delta n=1$.

Other quantities of interest are the (local) energy and momentum of the evolving STW as functions of time. To find these, we consider sample times $\tau_i=i \Delta t$, where $\Delta t=0.02$. At each time $t=\tau_i$, we determine the particle at which the strain reached its minimal value and compute the energy and momentum of a portion of the chain centered at this particle. The length of the portion, which is the same for each sample time $\tau_i$, is chosen so that the main body of the wave was included in it, which we took to be when the strain was of $O(10^{-4})$ at the ends. Typically, including $125$ particles is sufficient. Notice that while the total energy $H$ and momentum $P$ of the lattice remain conserved over the dynamical evolution (up to the relative error of $O(10^{-12})$ in the simulations), the localized energy and momentum portions associated with the wave may vary over time, especially in the scenario of the dynamical evolution of a spectrally unstable wave. In that light, these diagnostics are quite suitable for detecting the potential transformations of STWs as a result of their instability.

\section{Unstable dynamics in the $N$-region}
\label{sec:Nregion}
\begin{figure}[!htb]
\centering
\subfloat[]{\includegraphics[width=0.5\textwidth]{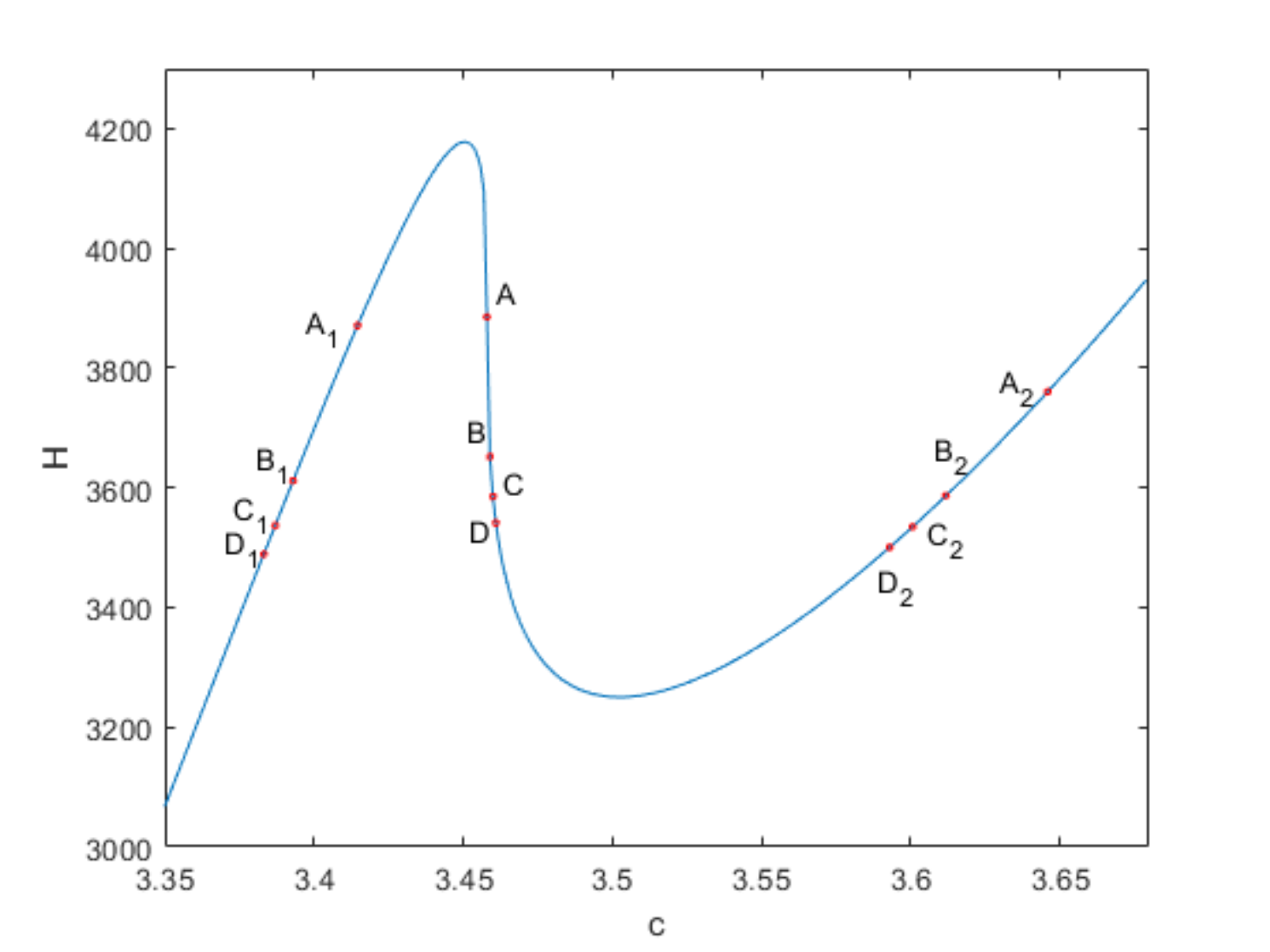}}
\subfloat[]{\includegraphics[width=0.5\textwidth]{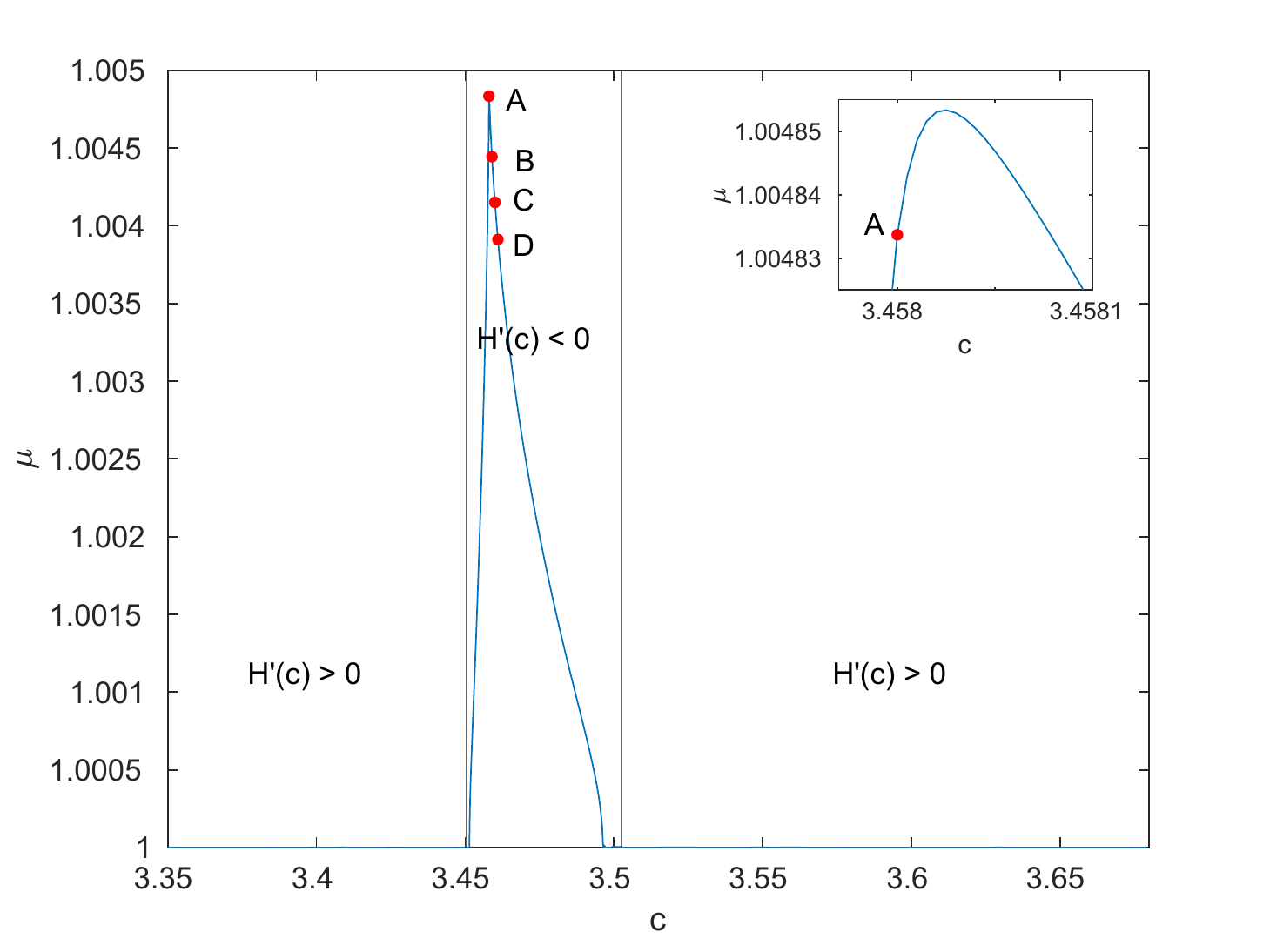}}
\caption{\footnotesize (a) Energy $H$ and (b) maximal real Floquet multiplier $\mu$ as functions of velocity $c$ of the STWs at $(\alpha,J)=(0.165,0.1)$. Unstable waves where $\mu>1$ correspond to the decreasing portion ($H'(c)<0$). Points $A$, $B$, $C$, and $D$ correspond to the velocities of the tested unstable waves, and points $A_1$, $A_2$, $B_1$, $B_2$, $C_1$, $C_2$, $D_1$, and $D_2$  mark the corresponding final velocities of the stable waves that the perturbed unstable STWs have evolved into, depending on the sign of the perturbation. Inset in (b) shows the enlarged view around the maximum.}
\label{fig:Energy}
\end{figure}

We start by investigating the unstable dynamics of STWs in the $N$ region. While multiple simulations in different regimes have been conducted, we present below only the results for $\alpha=0.165$, $J=0.1$ that are representative of the instability patterns observed in this parameter region. The corresponding $H$ and maximal real Floquet multiplier $\mu$ as functions of $c$ are shown in Fig.~\ref{fig:Energy}.

Due to translational invariance, the system always has a pair of unit Floquet multipliers, which are the maximal real multipliers in the velocity intervals corresponding to increasing energy ($H'(c)>0$). These velocity intervals apparently correspond to linearly stable STWs, although mild spurious oscillatory instabilities associated with complex Floquet multipliers slightly outside the unit circle may be present in this regime due to numerical artifacts that diminish as $L$ is increased \cite{Xu18}. As the first stability threshold is crossed, a symmetric pair of imaginary eigenvalues $\lambda$ collides at zero and reemerges on the real axis. Equivalently, a pair of multipliers sliding along the unit circle results in collision at the point $(1,0)$ of the
unit circle and reemerges on the real axis as a symmetric pair, with maximal real multiplier $\mu$ now exceeding one (and the second multiplier of the pair
now being inside the circle with a value of $1/\mu$), so that the corresponding STWs are unstable. The magnitude of $\mu$ increases, reaches a maximum value and then decreases again to one when the second stability threshold is crossed. It should be noted that in the numerical computations $H'(c)$ is slightly below zero at the two stability thresholds. As noted in \cite{Xu18}, this is an artifact of the finite length $L$ of the chain, and $H'(c)$ approaches zero at the threshold when $L$ is increased.

\begin{figure}[!htb]
\centering
\psfig{figure=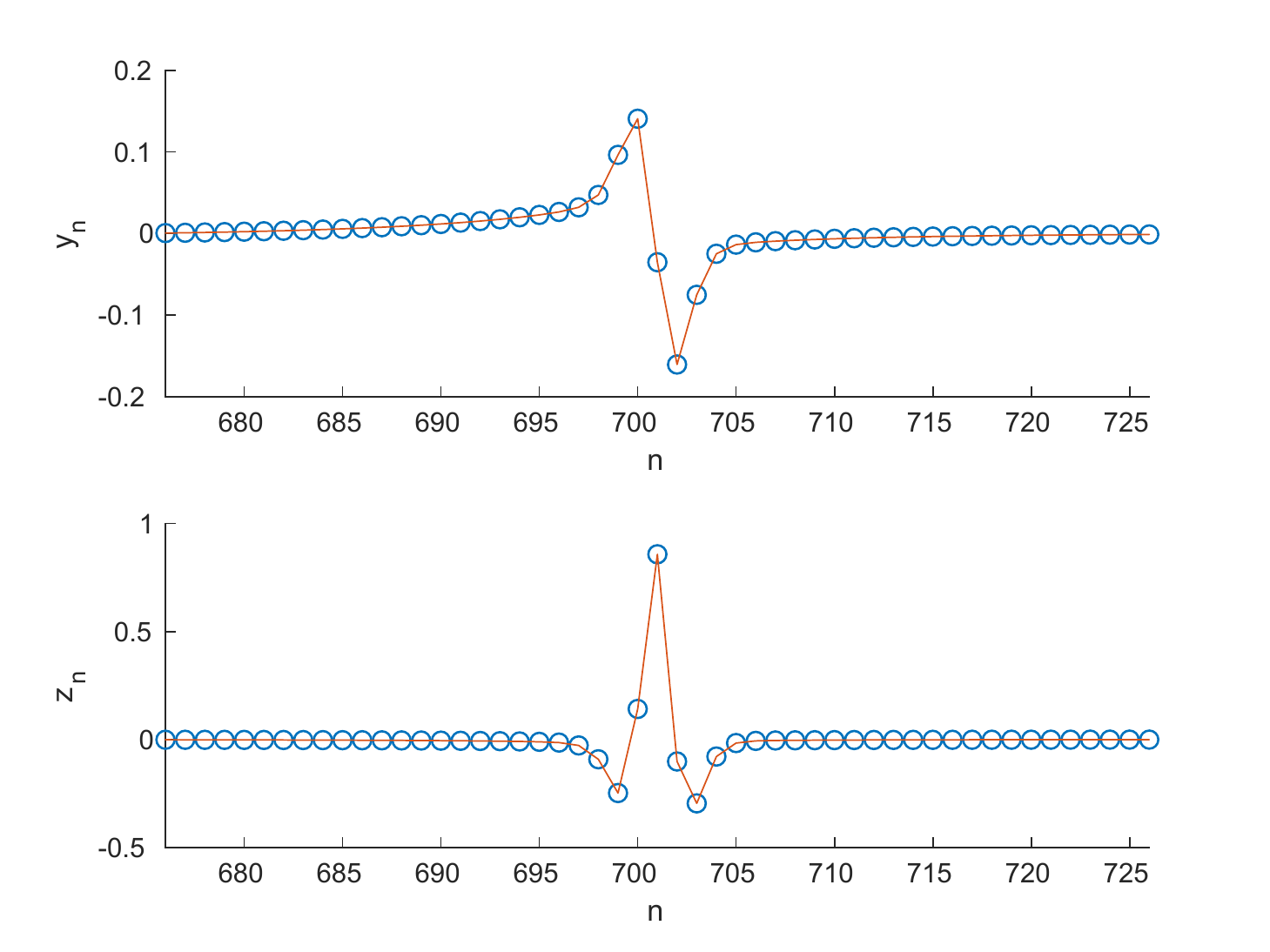,width=0.6\textwidth}
\caption{\footnotesize Eigenmode of an unstable STW with $c=3.458$, $(\alpha,J)=(0.165,0.1)$ corresponding to the Floquet multiplier $\mu=1.0048$ that leads to the speeding up of the perturbed wave. Here $y_n$ corresponds to strain and $z_n$ to its time derivative. Reversing the sign of the perturbation results in slowing down of the perturbed wave. }
\label{fig:Eigenmodes}
\end{figure}

To investigate the consequences of the instability associated with $\mu>1$, we selected STWs with four different velocities inside the unstable interval and perturbed them along the corresponding eigenmodes, as described in Sec.~\ref{sec:numer_methods}. The simulations were run until a stable propagation pattern  emerged. In all simulations, the perturbed unstable wave eventually evolves into a stable STW with lower energy and either smaller or higher velocity, as shown in Fig.~\ref{fig:Energy}. We found that the size of the perturbation only affected the time it takes for the stable waveform to emerge but not the resulting wave itself. We also found that adding small random noise (of amplitude $10^{-4}$) to the initial perturbation did not significantly affect the results, i.e., for a given unstable initial waveform, the
dynamical evolution would apparently select a unique end state on the corresponding
stable branches.
A typical eigenmode used to initiate the instability
is shown in Fig.~\ref{fig:Eigenmodes}. We note that each normalized eigenmode is determined up to plus or minus sign, so to change a wave from speeding up to slowing down or vice versa it suffices to reverse the sign of $\epsilon$.

\begin{figure}[!htb]
\centering
\subfloat[]{\includegraphics[width=0.33\textwidth]{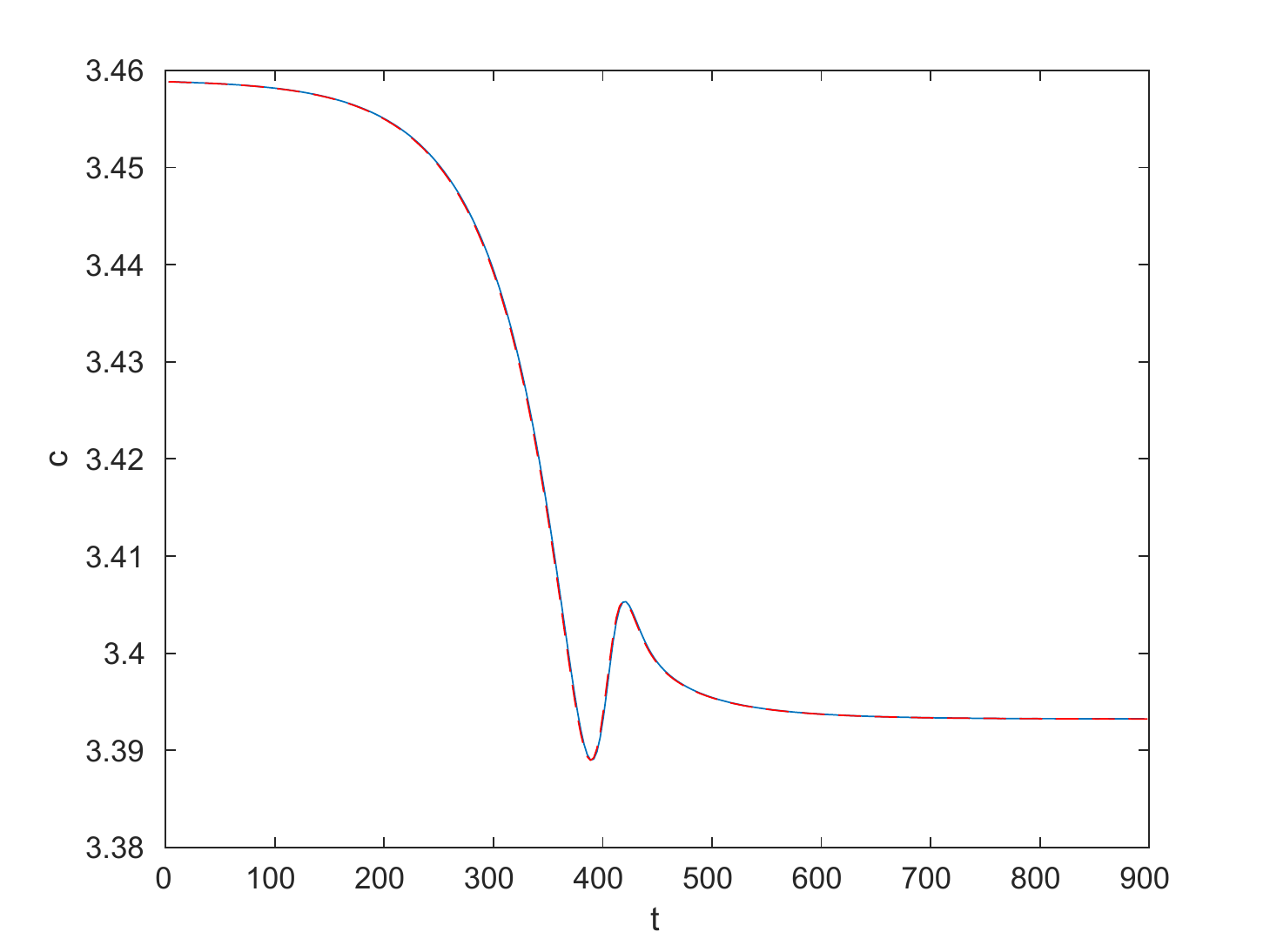}}\hfill
\subfloat[]{\includegraphics[width=0.33\textwidth]{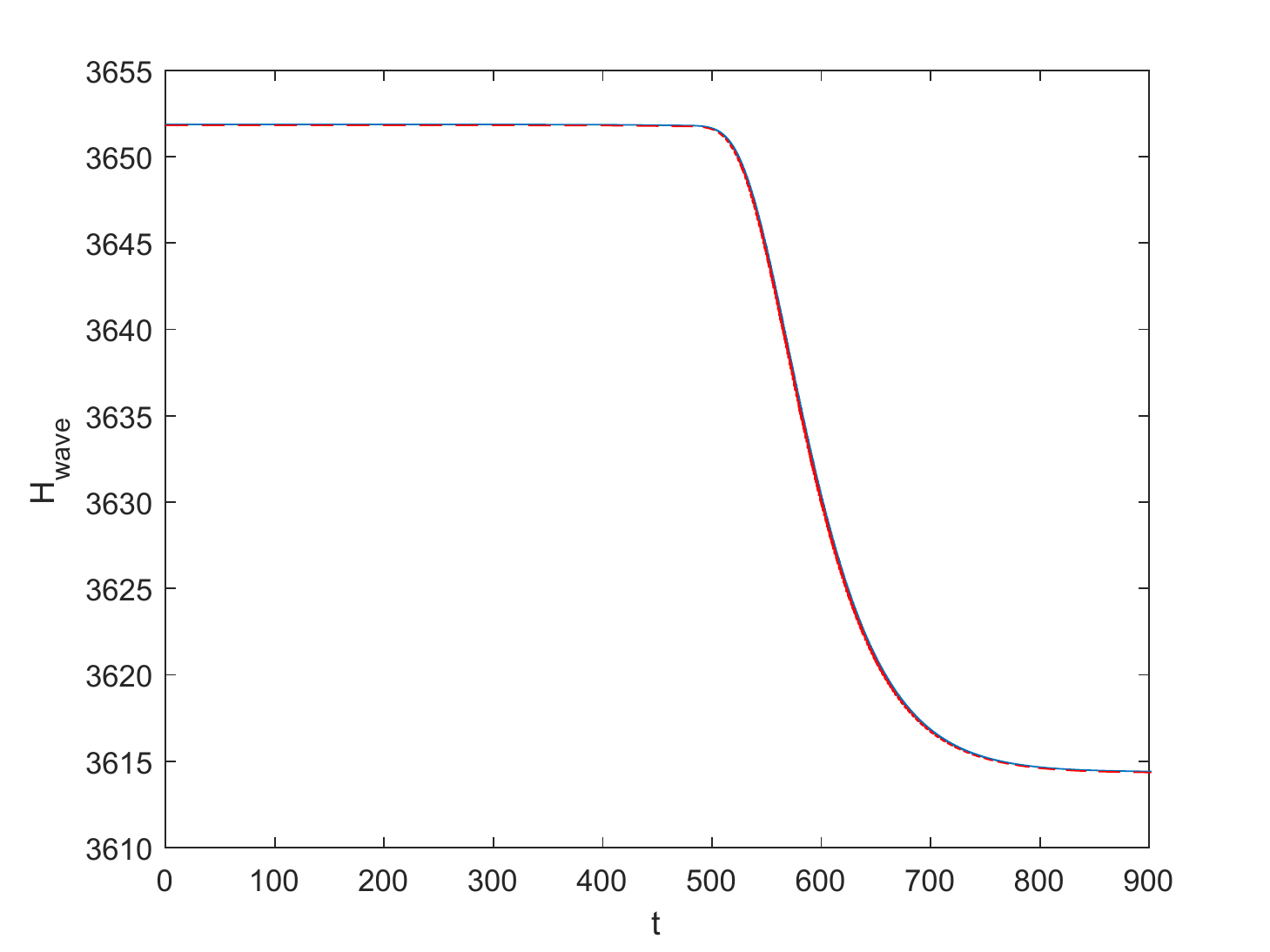}}\hfill
\subfloat[]{\includegraphics[width=0.33\textwidth]{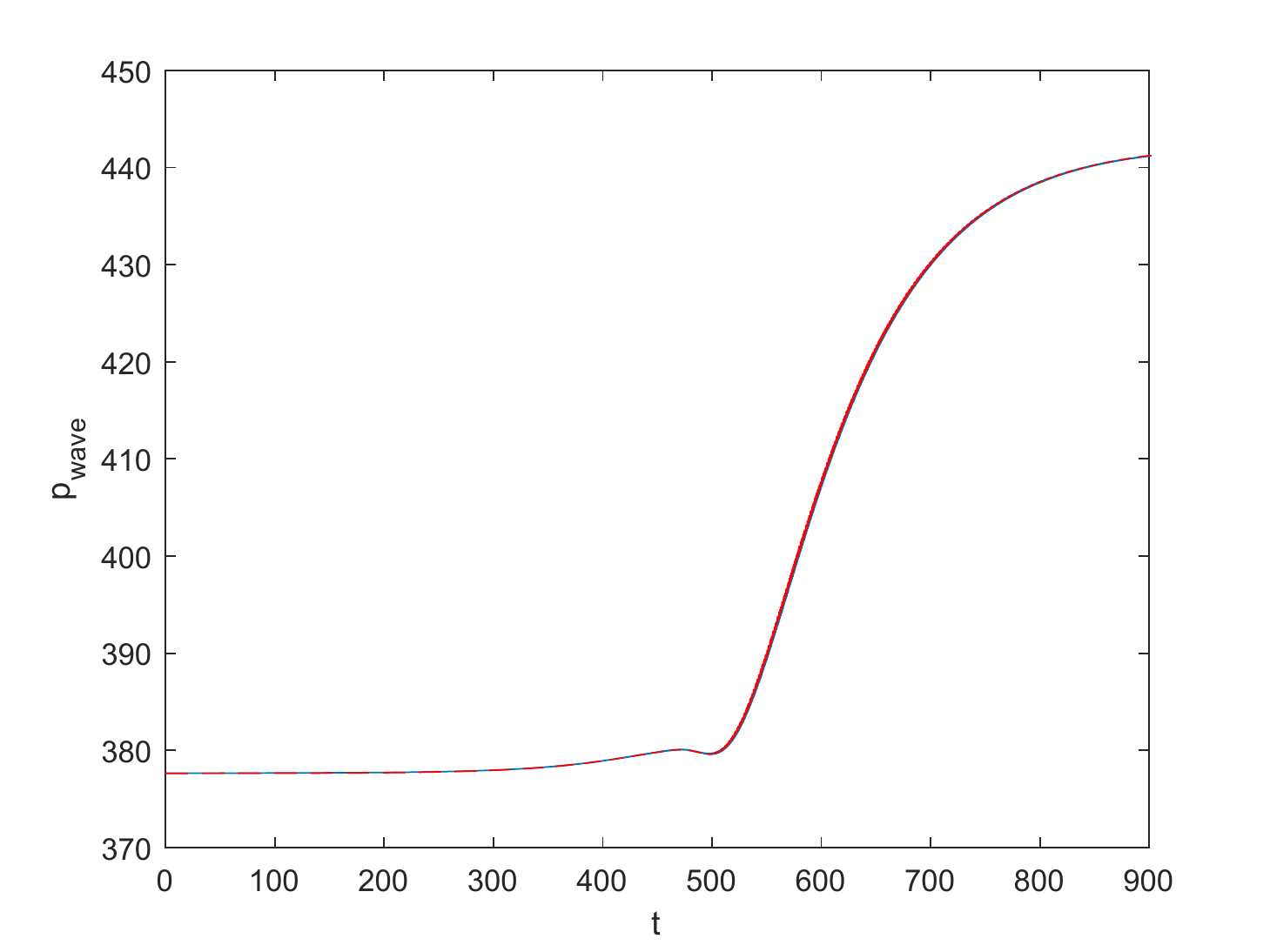}}
\caption{\footnotesize (a) Time evolution of the velocity of wave resulting from initial perturbation with $\epsilon=-0.25$ of the unstable STW with velocity  $3.459$ (point $B$ in Fig.~\ref{fig:Energy}) at $(\alpha,J)=(0.165,0.1)$. The velocity evolution is non-monotone: it initially decreases, then increases over a small time interval and then decreases again to the value $3.3932$ (point $B_1$ in Fig.~\ref{fig:Energy}) towards the end of the simulation. (b) Time evolution of the energy of the STW. (c) Time evolution of the momentum of the STW. The red dashed lines show the evolution with small-amplitude random noise added to the initial perturbation, while the solid blue lines correspond to the simulations without the additional noise.}
\label{fig:slow_cH}
\end{figure}

\begin{figure}[!htb]
\centering
\subfloat[]{\psfig{figure=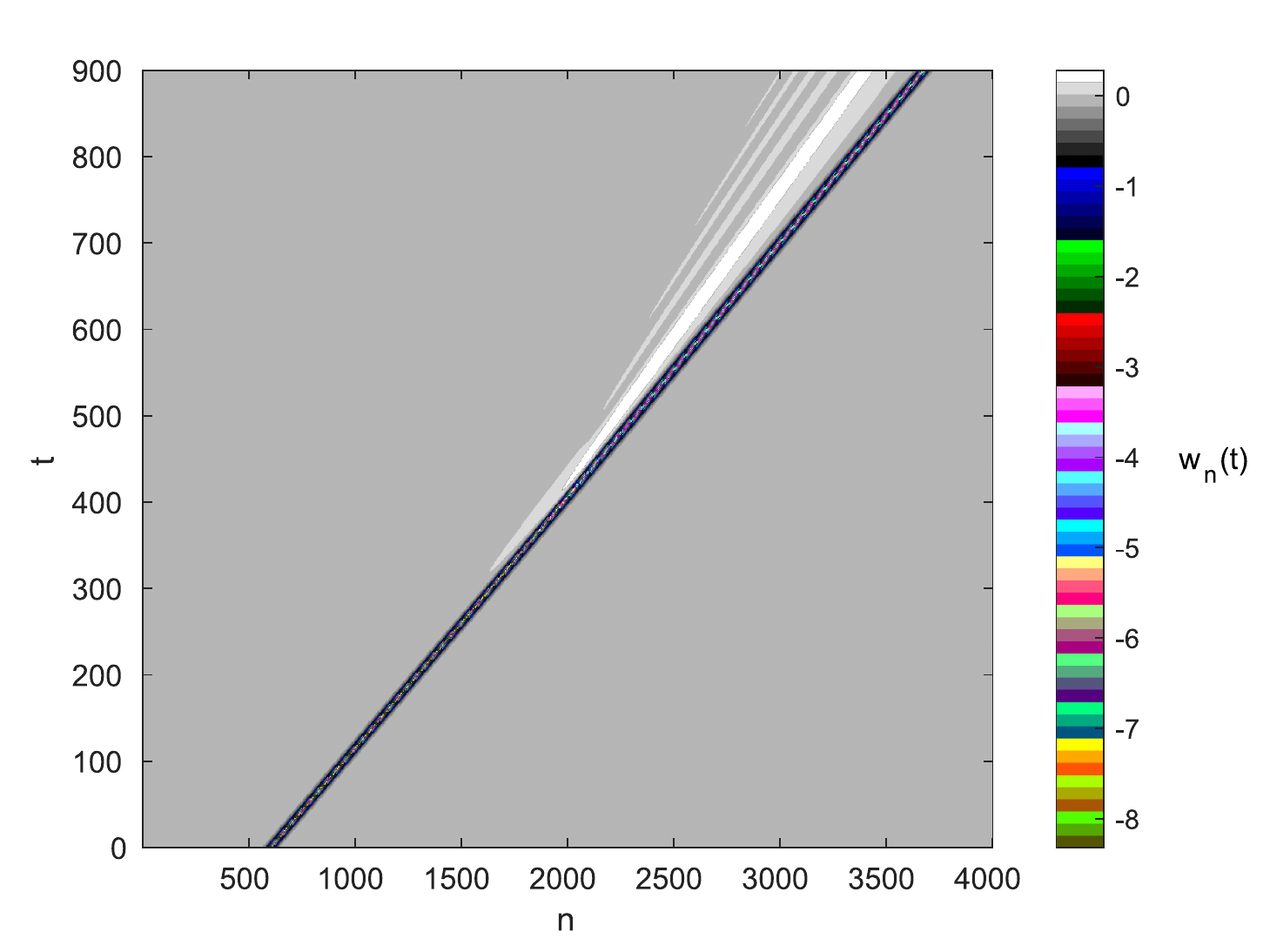,width=0.5\textwidth}}
\subfloat[]{\psfig{figure=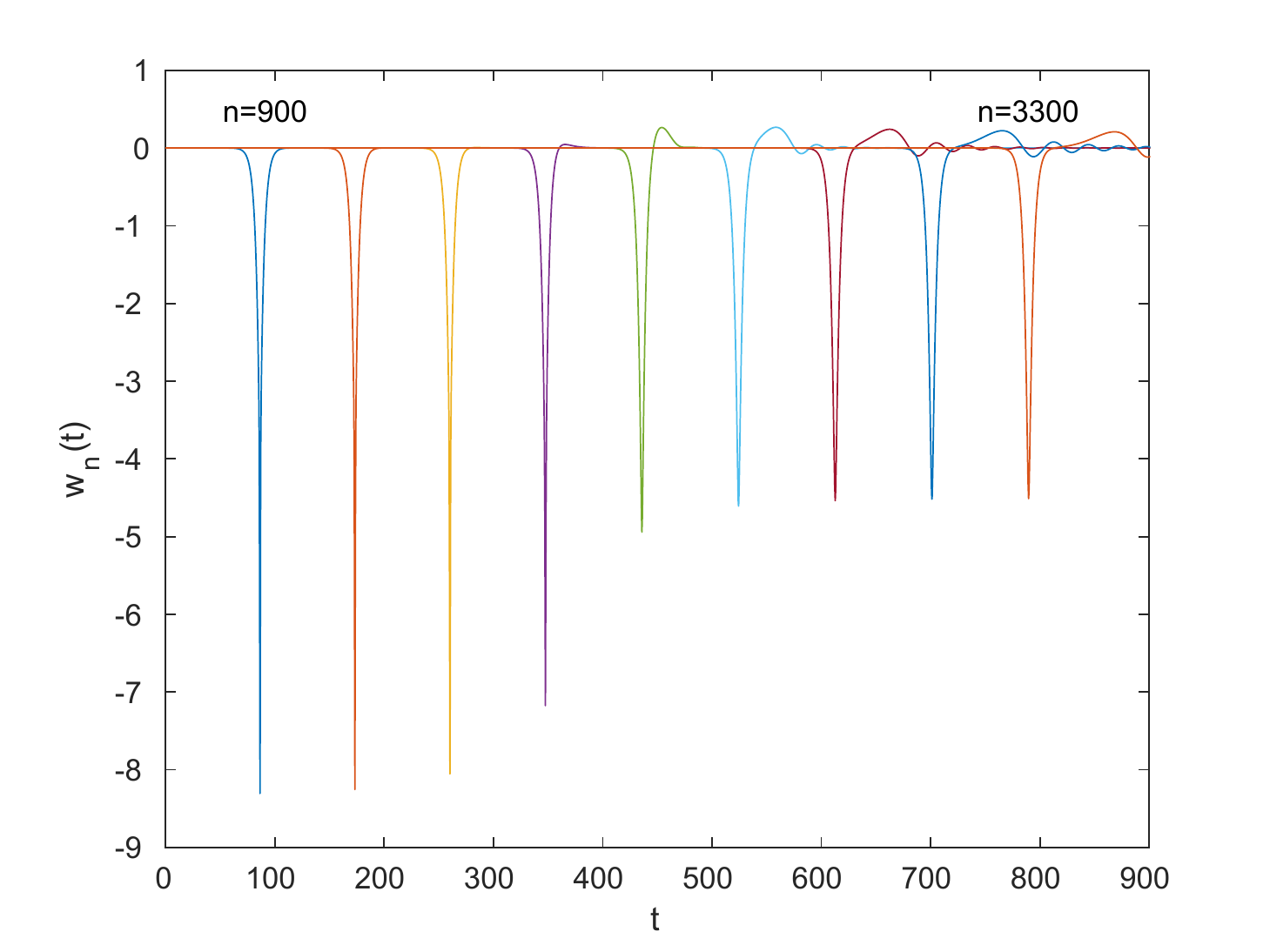,width=0.5\textwidth}}
\caption{\footnotesize  (a) Space-time and (b) time evolution of $w_n(t)$ at fixed $n$ during the transition from $B$ to $B_1$ shown in Fig.~\ref{fig:slow_cH}. A primarily tensile dispersive shock wave is expelled by the main waveform as it slows down. Here $n_0=701$, and the selected values of $n$ are spaced $300$ units apart in (b).}
\label{fig:slow_evol}
\end{figure}

\begin{figure}[!htb]
\centering
\subfloat[]{\psfig{figure=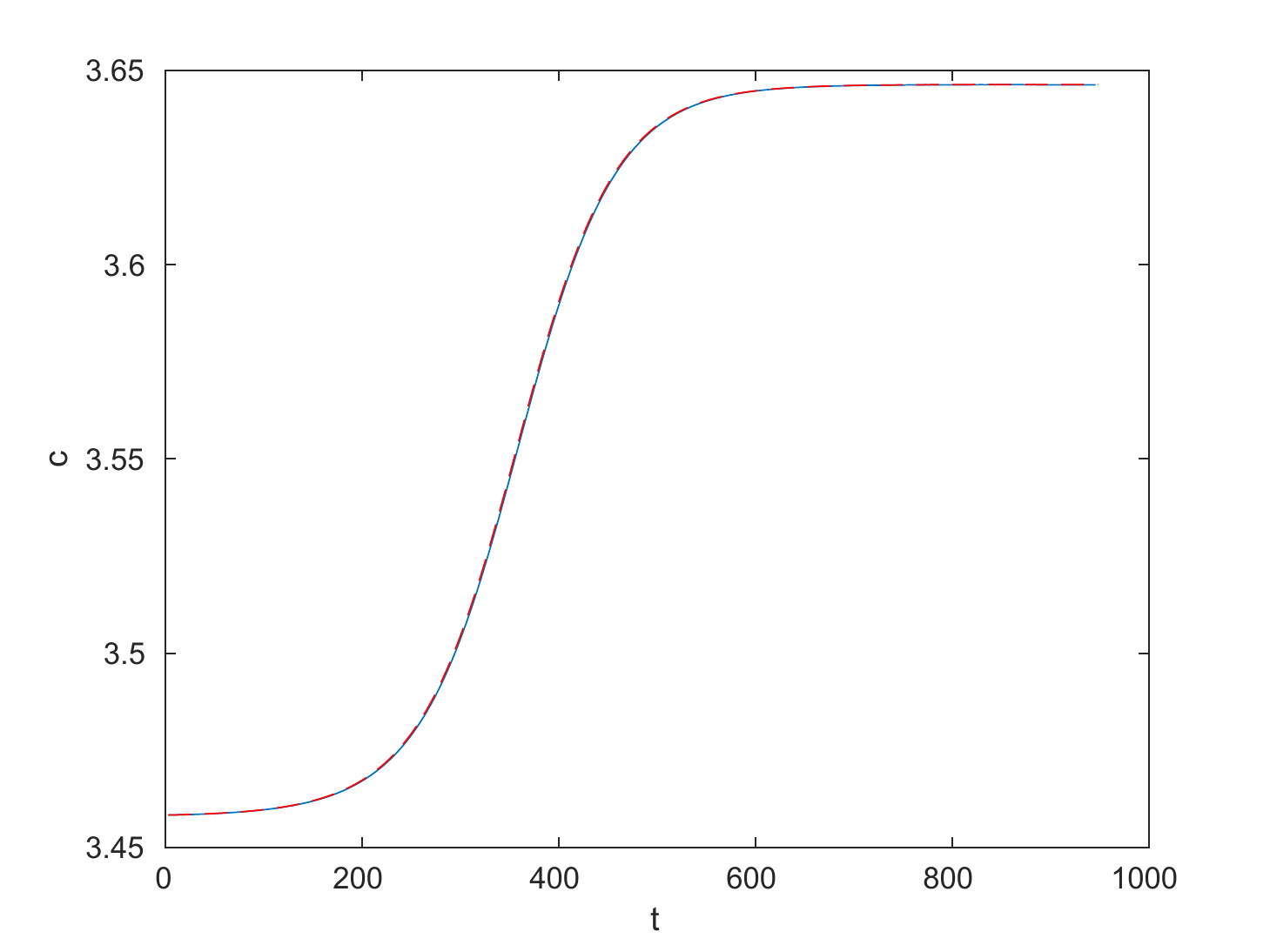,width=0.33\textwidth}}\hfill
\subfloat[]{\psfig{figure=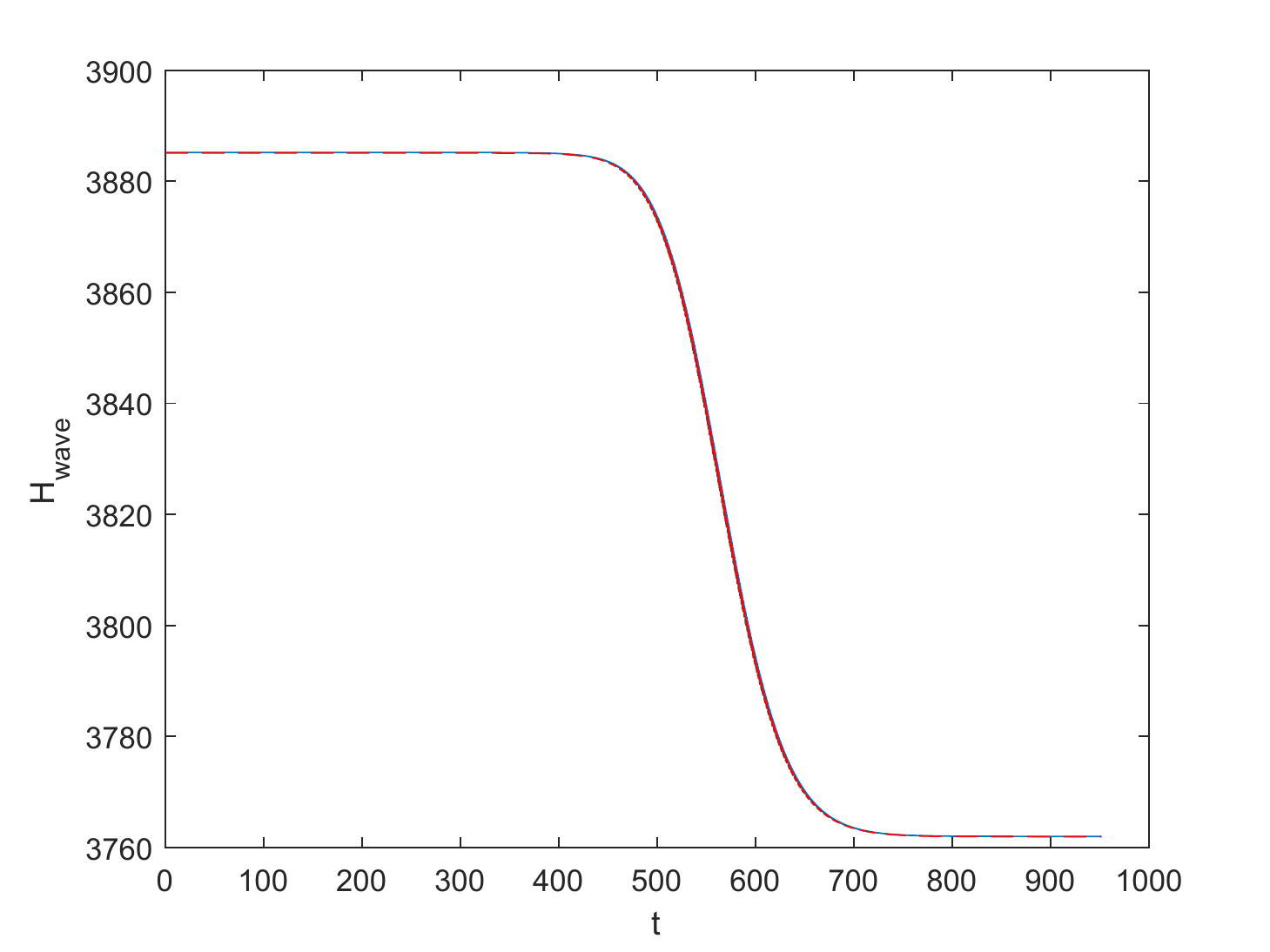,width=0.33\textwidth}}\hfill
\subfloat[]{\psfig{figure=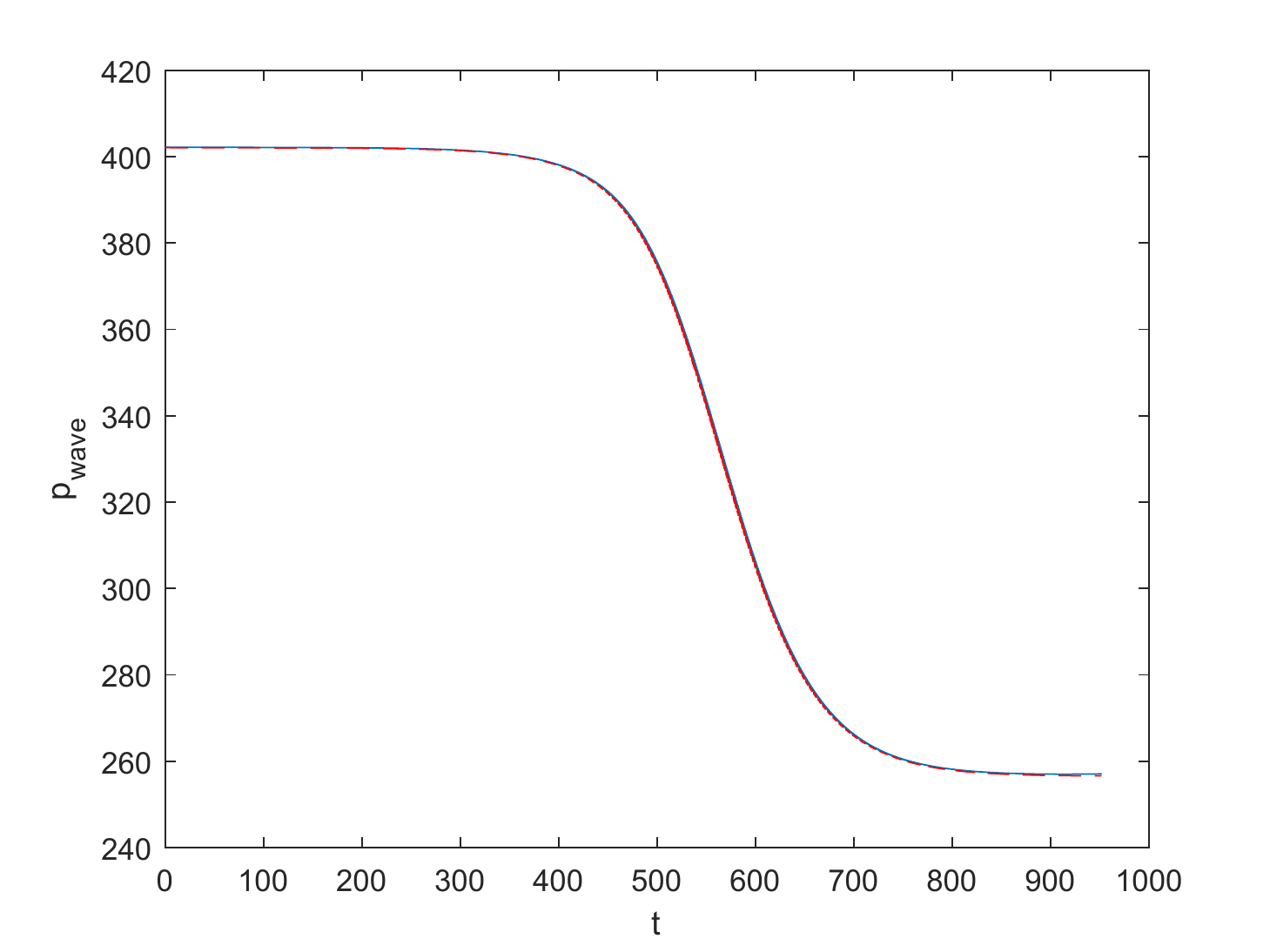,width=0.33\textwidth}}
\caption{\footnotesize (a) Time evolution of the velocity of wave resulting from initial perturbation with $\epsilon=0.25$ of the unstable STW with velocity $3.458$ (point $A$ in Fig.~\ref{fig:Energy}) at $(\alpha,J)=(0.165,0.1)$. The velocity increases, approaching the value $3.6462$ (point $A_2$ in Fig.~\ref{fig:Energy}) towards the end of the simulation. (b) Time evolution of the energy of the STW. (c) Time evolution of the momentum of the STW. The red dashed lines show the evolution with small-amplitude random noise added to the initial perturbation, while the solid blue lines correspond to the simulations without the additional noise.}
\label{fig:fast_cHp}
\end{figure}

Representative examples of velocity, energy and momentum evolution are shown in Fig.~\ref{fig:slow_cH} and Fig.~\ref{fig:fast_cHp}. We observed that when the velocity of the perturbed unstable wave eventually decreases, the wave expels a small-amplitude dispersive shock wave, as can be seen in Fig.~\ref{fig:slow_evol}. As shown in Fig.~\ref{fig:slow_cH}(a), the velocity evolution in this case is nonmonotone: after initially decreasing, it briefly increases then decreases again to the final value. These velocity oscillations take place right around the
time the dispersive wave formation becomes visible in the space-time plot shown in Fig.~\ref{fig:slow_evol}(a). Once this trailing dispersive wave detaches from the primary supersonic STW, the latter settles towards its final velocity. Note that while the energy of the wave decreases during this evolution, its momentum increases, with the total momentum of the system kept constant due to the negative contribution of the dispersive wave.

\begin{figure}[!htb]
\centering
\subfloat[]{\psfig{figure=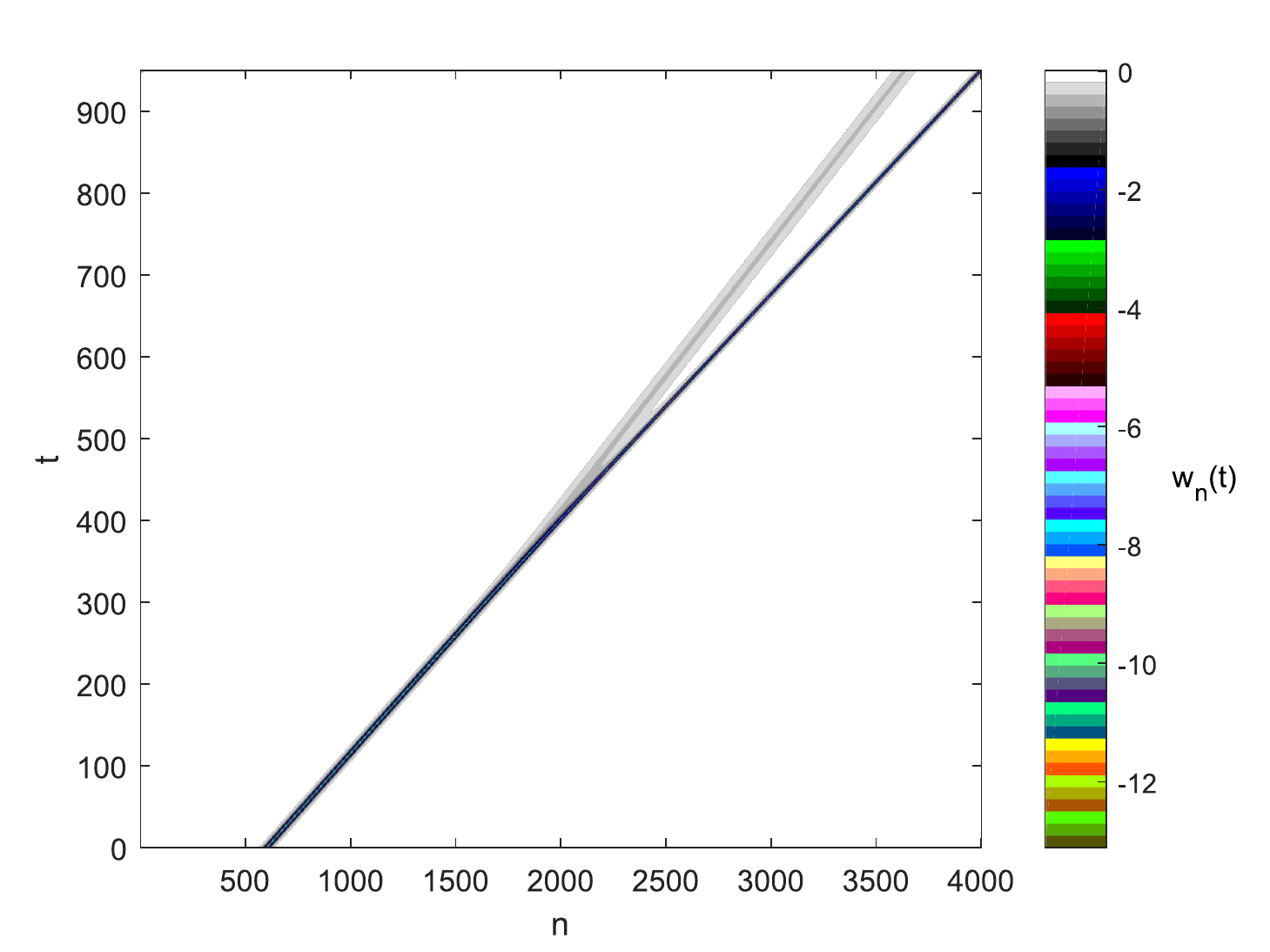,width=0.5\textwidth}}
\subfloat[]{\psfig{figure=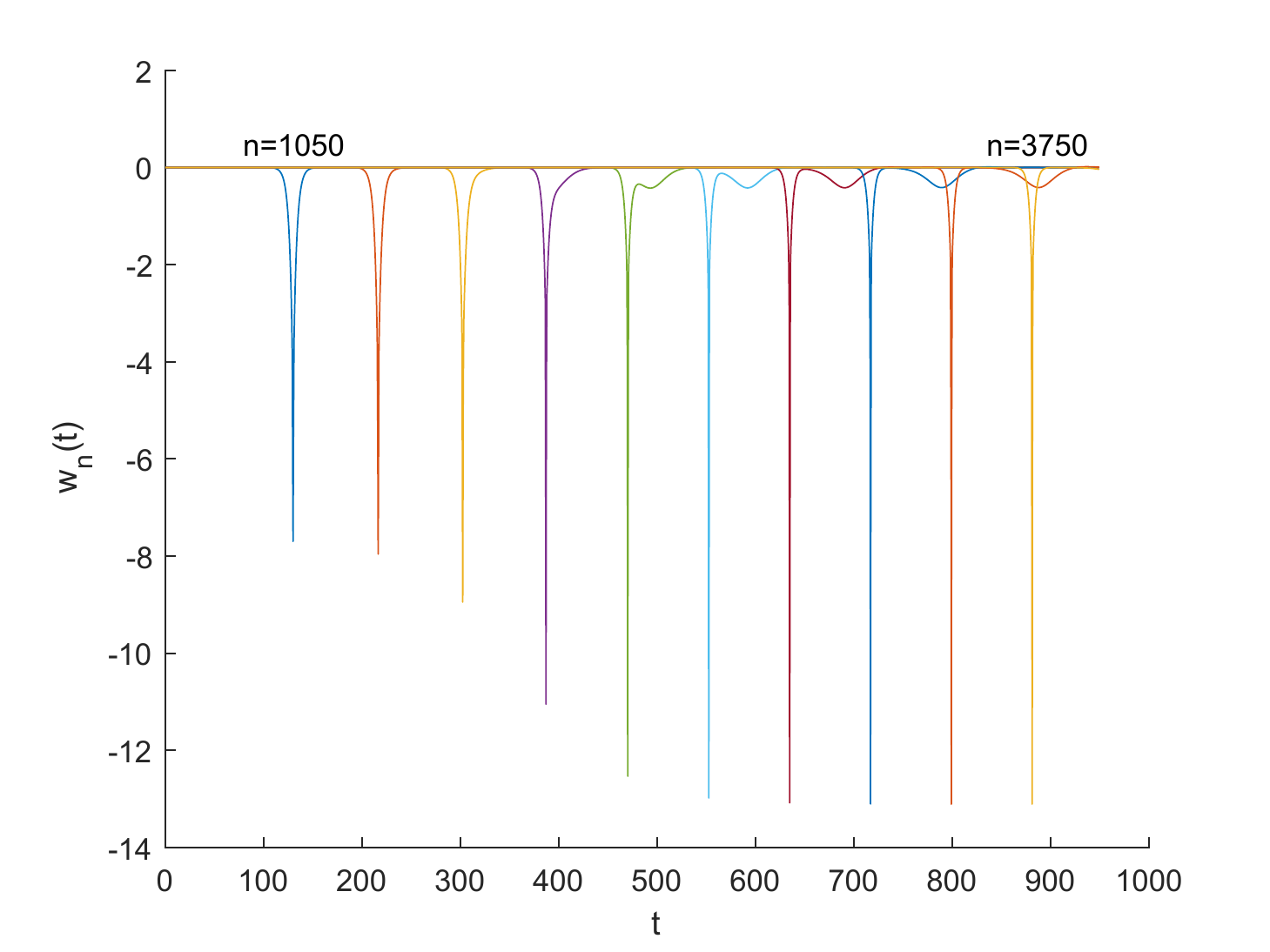,width=0.5\textwidth}}
\caption{\footnotesize (a) Space-time and (b) time evolution of $w_n(t)$ at fixed $n$ during the transition from $A$ to $A_2$ shown in Fig.~\ref{fig:fast_cHp}. A compressive small-amplitude STW, trailed by small amplitude oscillations, forms behind the main nonlinear waveform and eventually separates from it as the main wave increases its velocity. Here $n_0=701$, and the selected values of $n$ are spaced $300$ units apart in (b).}
\label{fig:fast_evol}
\end{figure}

The dynamics is quite different when the velocity of the perturbed unstable wave increases (see Fig.~\ref{fig:fast_cHp}). In this case, a small-amplitude STW, trailed by small-amplitude oscillations, forms behind the main waveform and eventually separates from it since it travels with smaller velocity; see Fig.~\ref{fig:fast_evol}. In this case the momentum of the primary wave decreases during the evolution due to the positive momentum of the slower wave.

\section{Results for the $Z$-region}
\label{sec:Zregion}
\begin{figure}[!htb]
\centering
\subfloat[]{\includegraphics[width=0.5\textwidth]{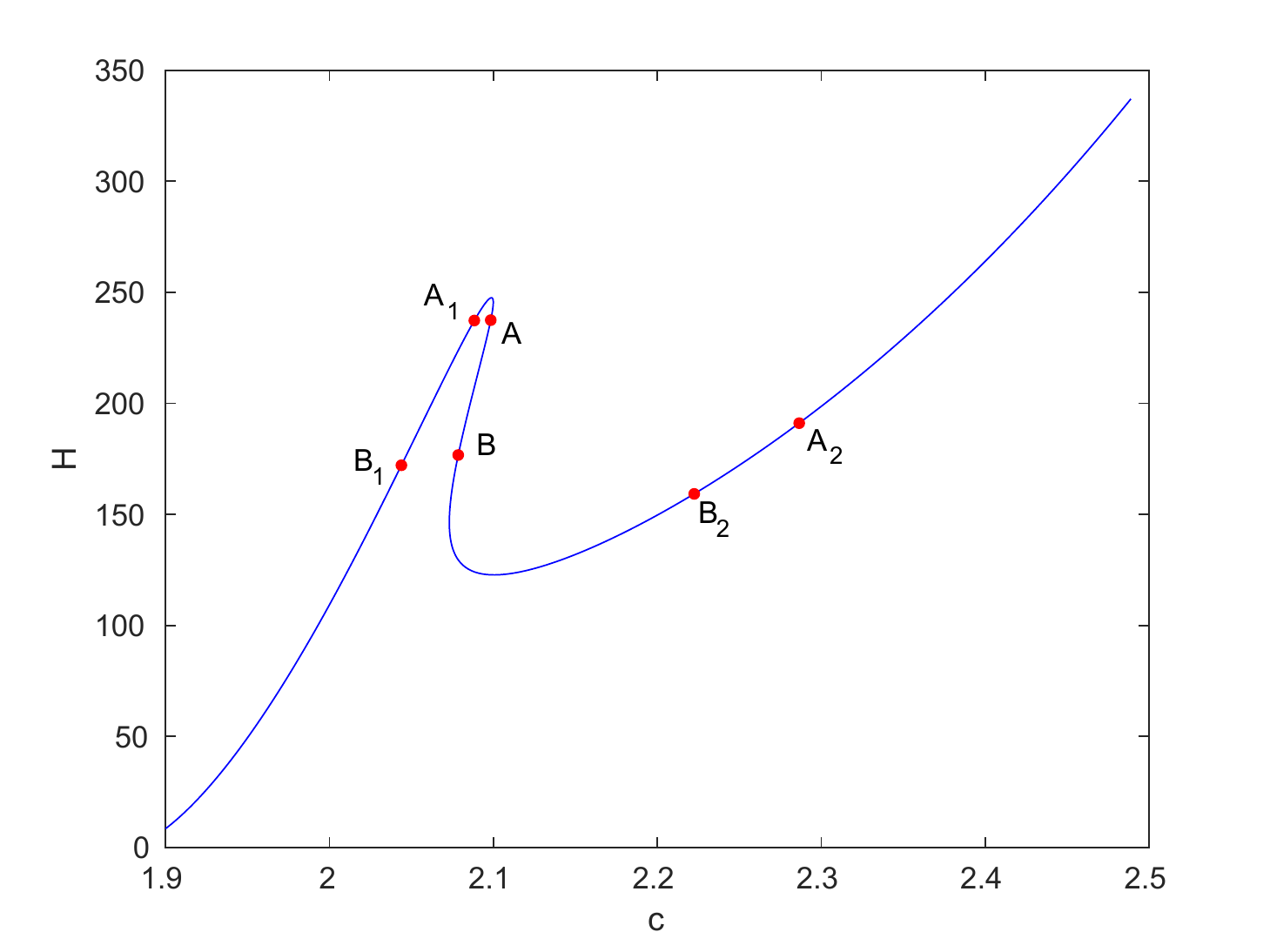}}\hfill
\subfloat[]{\includegraphics[width=0.5\textwidth]{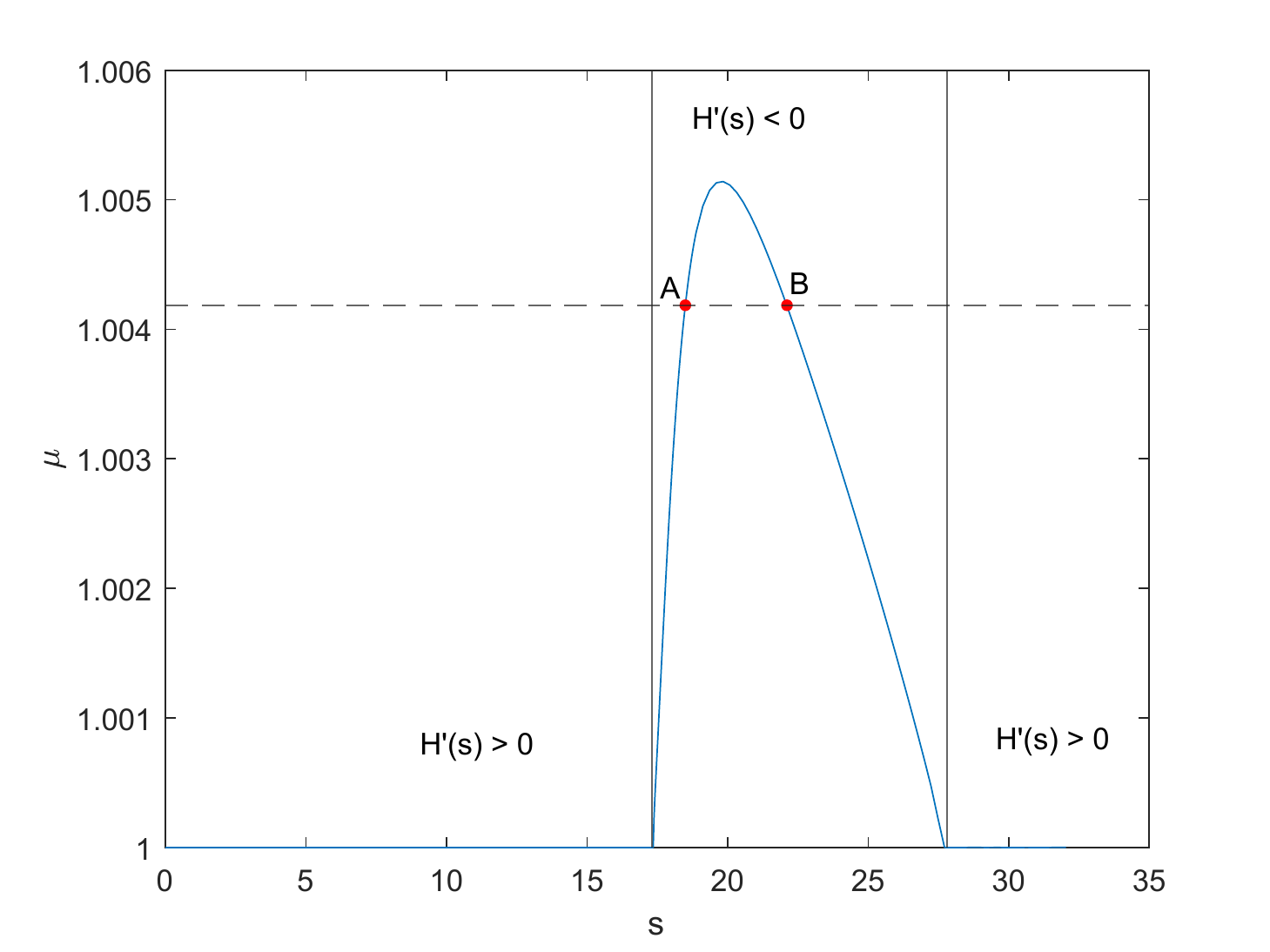}}
\caption{\footnotesize (a) Energy $H$ versus velocity $c$ of STWs at $(\alpha,J)=(0.1,0.012)$. Points $A$ and $B$ correspond to the energy and velocity of the tested waves, and points $A_1$, $A_2$, $B_1$, and $B_2$ mark the corresponding final velocities and energies of the stable waves the perturbed unstable STWs have evolved into. (b) Maximal real Floquet multiplier $\mu$ as a function of the parameter $s$. The solid vertical lines indicate the values of $s$ where $H'(s)=0$. The dashed horizontal line indicates the value $\mu=1.0042$. In both figures, point $A$ corresponds to the STW with velocity $2.0984$ and point $B$ corresponds to the STW with velocity $2.0785$.}
\label{fig:Zregion}
\end{figure}

We now consider the $Z$-region.
Recall that in this parameter region the function $H(c)$ becomes multivalued in a certain velocity interval. Using the pseudo-arclength algorithm, as described in Sec.~\ref{sec:numer_methods}, we computed such curves and analyzed
the linear stability of the corresponding STWs for various parameter values in the region. Below we just describe the representative case $\alpha=0.1$, $J=0.012$. The energy-velocity plot for these parameter values is shown in Fig.~\ref{fig:Zregion}(a).
Along the curve $c=c(s)$ and $H=H(s)$, and each of these is a nonmonotone up-down-up function, so that both $H'(s)$ and $c'(s)$ change sign twice, i.e.,
$H(c)$ is triple-valued within a relevant  interval of Fig.~\ref{fig:Zregion}(a). However, the changes in monotonicity of $H(s)$ and $c(s)$ do not take place simultaneously. Specifically, the first sign change for $H'(s)$, from positive to negative, occurs slightly before $c(s)$ starts decreasing, and $c'(s)$ changes its sign back to positive prior to $H'(s)$. Thus we have $c'(s)>0$ at both values of $s$ where $H'(s)$ crosses zero.

As discussed in the Appendix, each threshold value of $s$ where $H'(s)=0$ corresponds to a change in stability due to the increase in multiplicity of the zero eigenvalue of the operator associated with the linearized problem. The Hamiltonian nature of the problem implies that at the threshold value a symmetric pair of eigenvalues is meeting at the origin and is emerging on the real axis as $\pm \lambda$, $\lambda>0$, as the wave becomes unstable, so that a real Floquet multiplier $\mu=\exp(\lambda/c)>1$ appears in the unstable regime.
To verify this for our numerically computed STWs, we plot in Fig.~\ref{fig:Zregion}(b) the maximal real Floquet multiplier $\mu$ as the function of $s$ for the obtained solutions. One can see that $\mu>1$ in the interval of $s$ that nearly coincides with the one where $H'(s)<0$ (similarly to the observations in the previous section, $H'(s)$ is slightly below zero at the threshold values due to the finite length of the computational domain, though this numerical artifact is not visible in Fig.~\ref{fig:Zregion}(b)). Thus, three STWs coexist for each $c$ in the velocity interval where $c'(s)<0$. Among these, the waves where $H'(s)<0$ are unstable. This always includes the intermediate-energy wave, in agreement with the numerical observations in \cite{MGM00}, but low-energy and high-energy waves also become unstable near the left and right ends of the velocity interval, respectively.

\begin{figure}[!htb]
\centering
\subfloat[]{\includegraphics[width=0.5\textwidth]{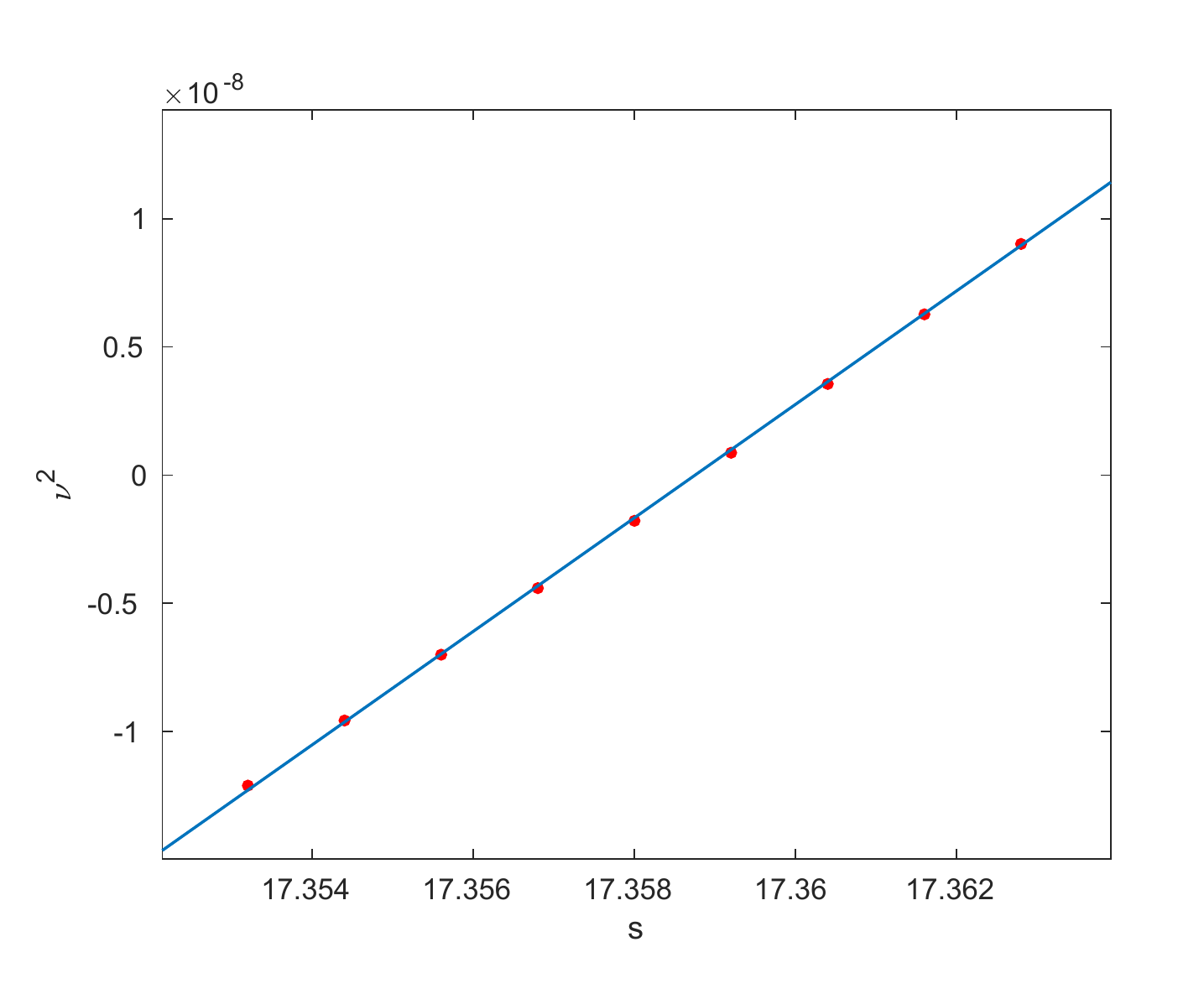}}
\subfloat[]{\includegraphics[width=0.5\textwidth]{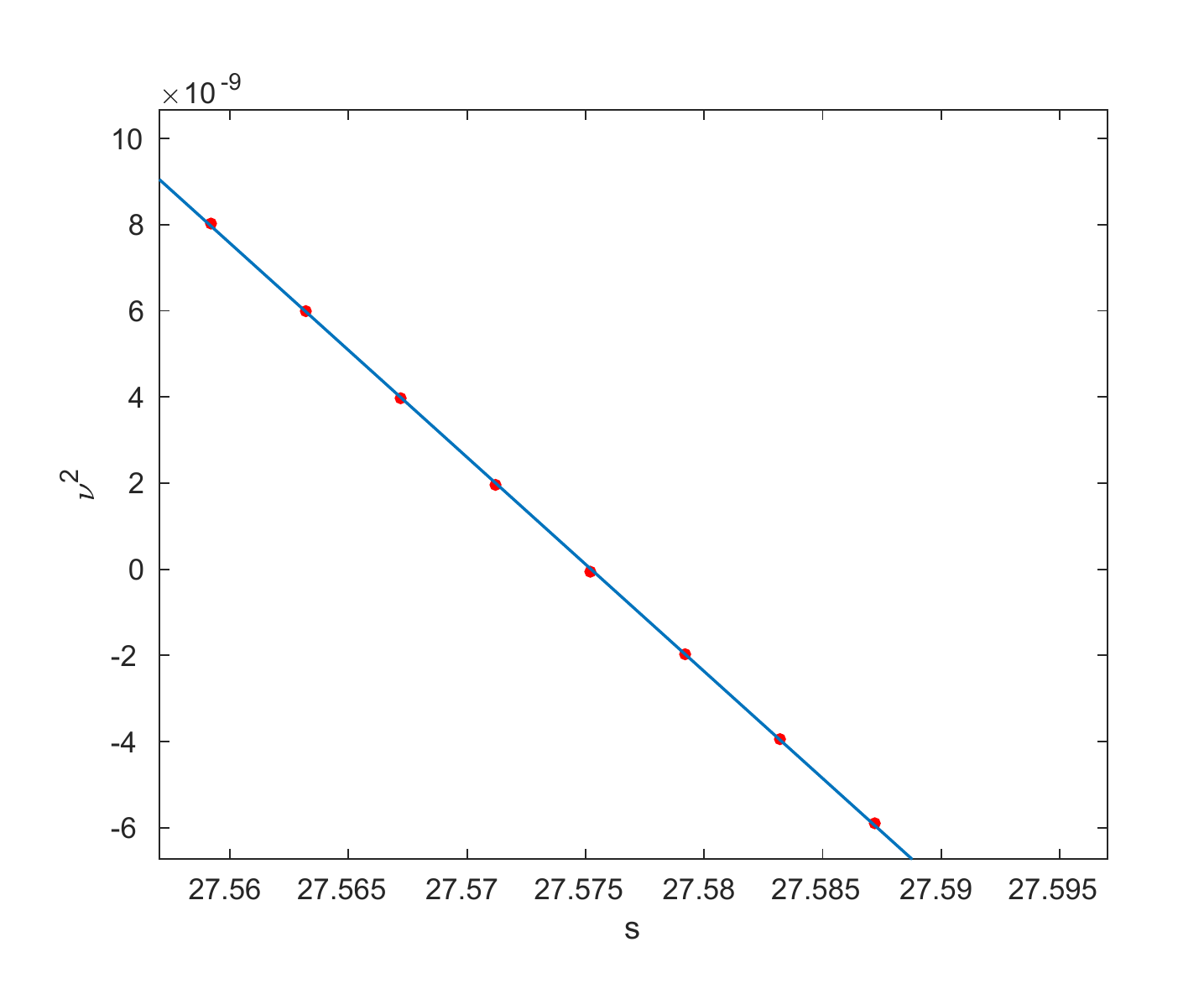}}
\caption{\footnotesize Squared rescaled near-zero eigenvalues $\nu(s)=\lambda(s)/c(s)=\ln(\mu(s))$ near (a) the maximum and (b) the minimum of $H(s$) at $(\alpha,J)=(0.1,0.012)$. The straight lines show the best linear fit in each case.}
\label{fig:splitting}
\end{figure}

The splitting of the zero eigenvalue and transition to instability near the maximum and minimum of $H(s)$ is illustrated in Fig.~\ref{fig:splitting}.
The plots show $\nu^2(s)$, where $\nu(s)=\lambda(s)/c(s)$ is a rescaled near-zero eigenvalue (note that $\nu=\ln(\mu)$, where $\mu$ is the corresponding Floquet multiplier near $1$). As the stability threshold is crossed into the unstable region in each case, a symmetric pair of purely imaginary eigenvalues ($\nu^2<0$) becomes a symmetric pair of real ones ($\nu^2>0$). The fact that $\nu^2 \sim s-s_0$ near each threshold $s_0$ is in agreement with the approximation \eqref{eq:split} derived in the Appendix.

We now examine the dynamical fate of unstable solutions. We consider two cases with velocities $2.0785$ and $2.0984$ that have the same Floquet multiplier $\mu=1.0042$, which corresponds to eigenvalues $\lambda=0.0087$ and $0.0088$, respectively. Similar to the previously discussed cases for the $N$-region, the waves either slow down after expelling a dispersive wave or speed up after expelling a small-amplitude solitary wave, depending on the sign of the perturbation $\epsilon$.

\begin{figure}[!htb]
\centering
\subfloat[]{\includegraphics[width=0.5\textwidth]{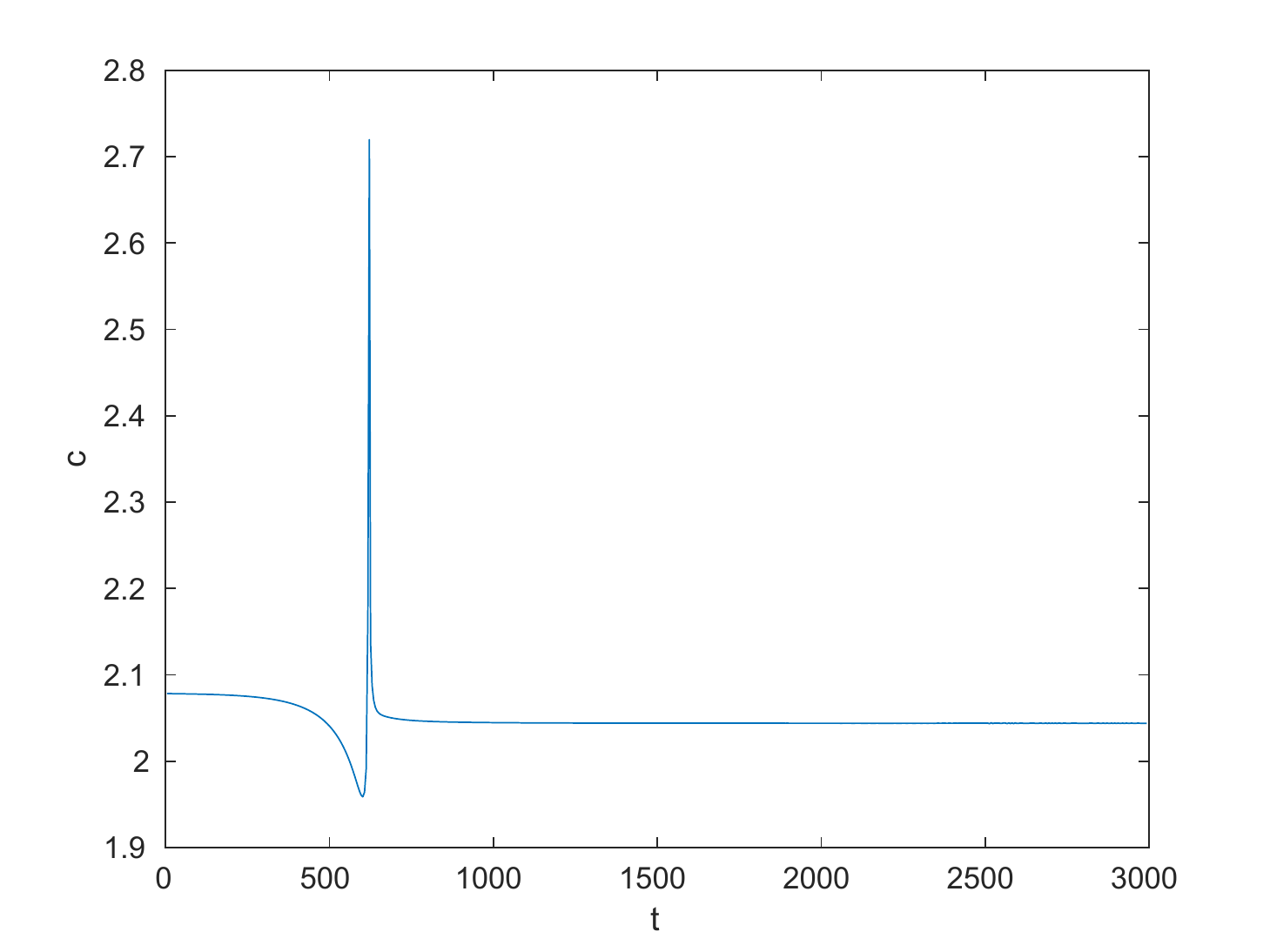}}
\subfloat[]{\includegraphics[width=0.5\textwidth]{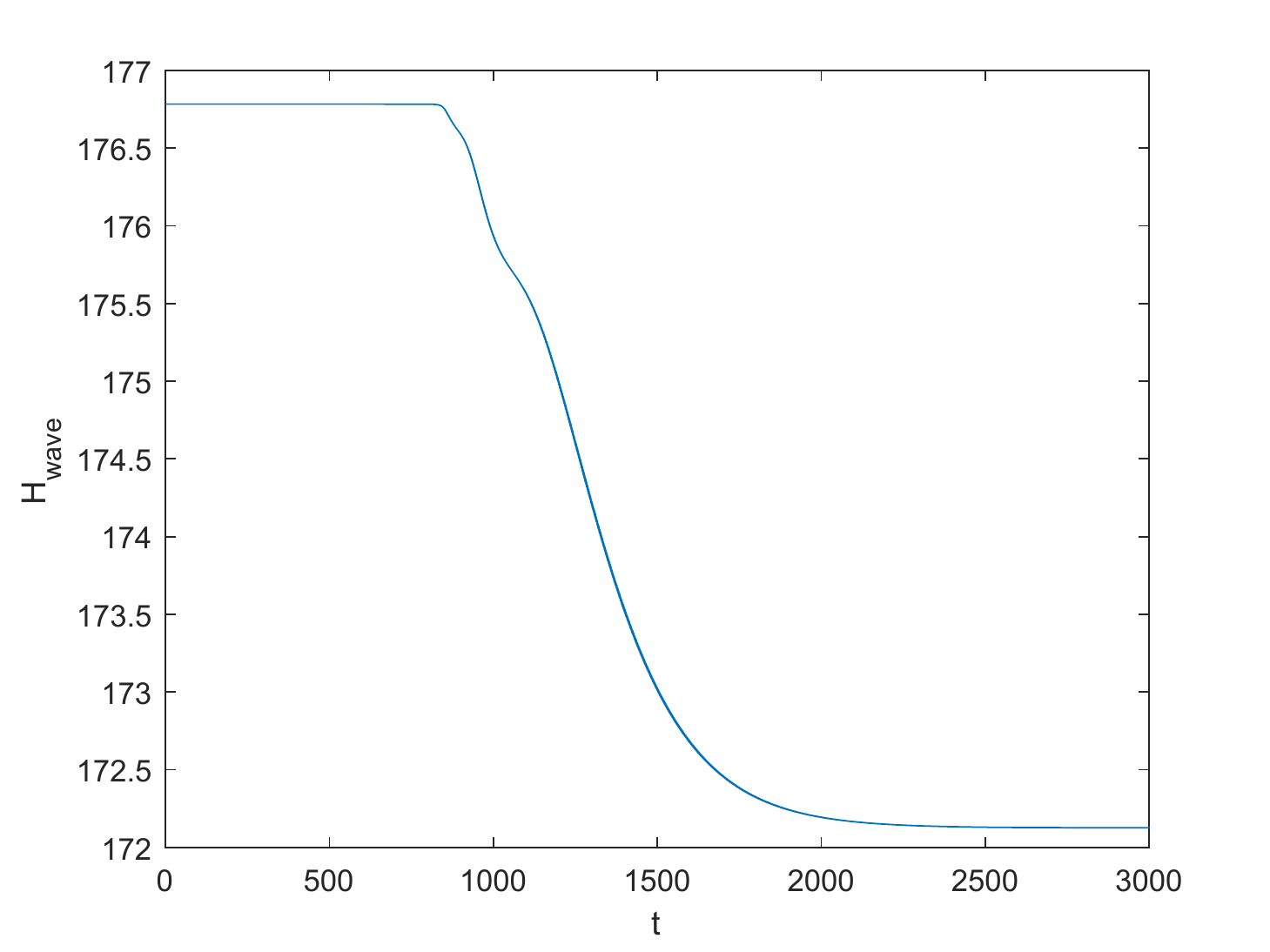}}
\caption{\footnotesize (a) Time evolution of the velocity of wave resulting from initial perturbation with $\epsilon=-0.25$ of the unstable STW with velocity $2.0785$ (point $B$ in Fig.~\ref{fig:Zregion}(a)) at $(\alpha,J)=(0.1,0.012)$. The final velocity is $2.0439$ (point $B_1$ in Fig.~\ref{fig:Zregion}(a)). (b) Time evolution of the energy of the STW.}
\label{fig:slow_cH_B}
\end{figure}

\begin{figure}[!htb]
\centering
\subfloat[]{\psfig{figure=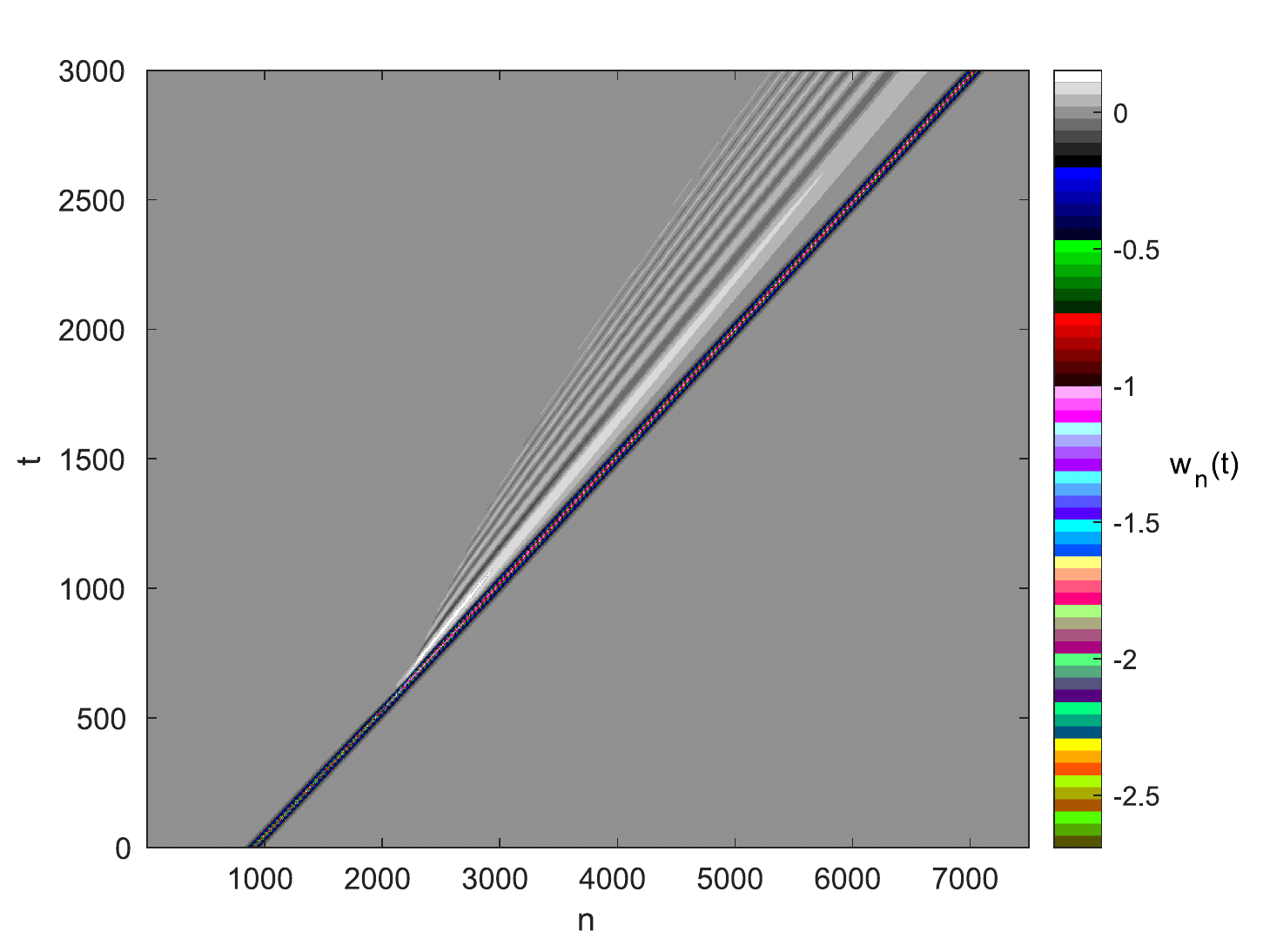,width=0.5\textwidth}}
\subfloat[]{\psfig{figure=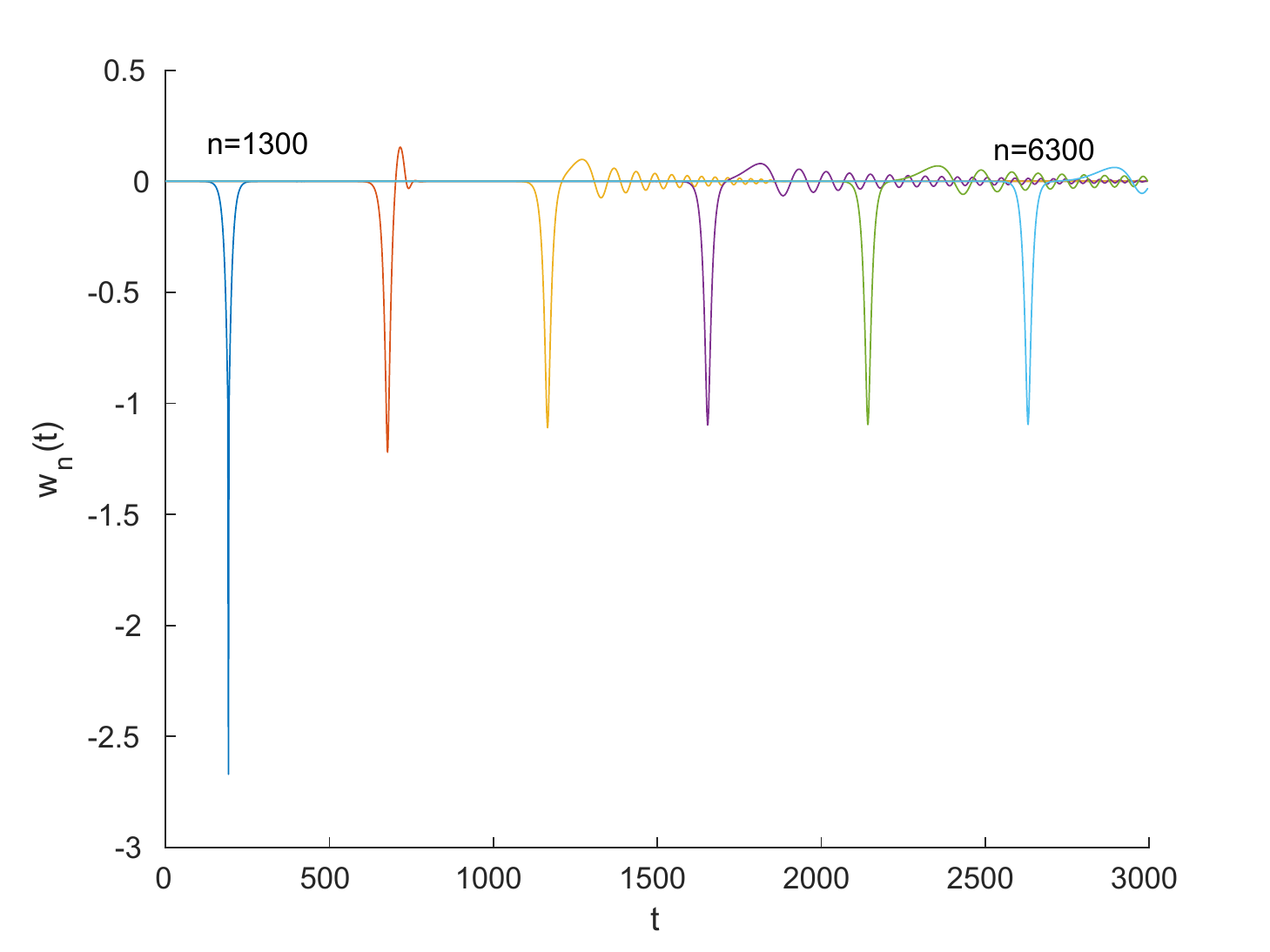,width=0.5\textwidth}}
\caption{\footnotesize  (a) Space-time and (b) time evolution of $w_n(t)$ at fixed $n$ during the transition from $B$ to $B_1$ shown in Fig.~\ref{fig:slow_cH_B}. A dispersive shock wave is expelled by the main waveform as it slows down. Here $n_0=901$, and the selected values of $n$ are spaced $1000$ units apart in (b).}
\label{fig:slow_evol_B}
\end{figure}

\begin{figure}[!htb]
\centering
\includegraphics[width=0.5\textwidth]{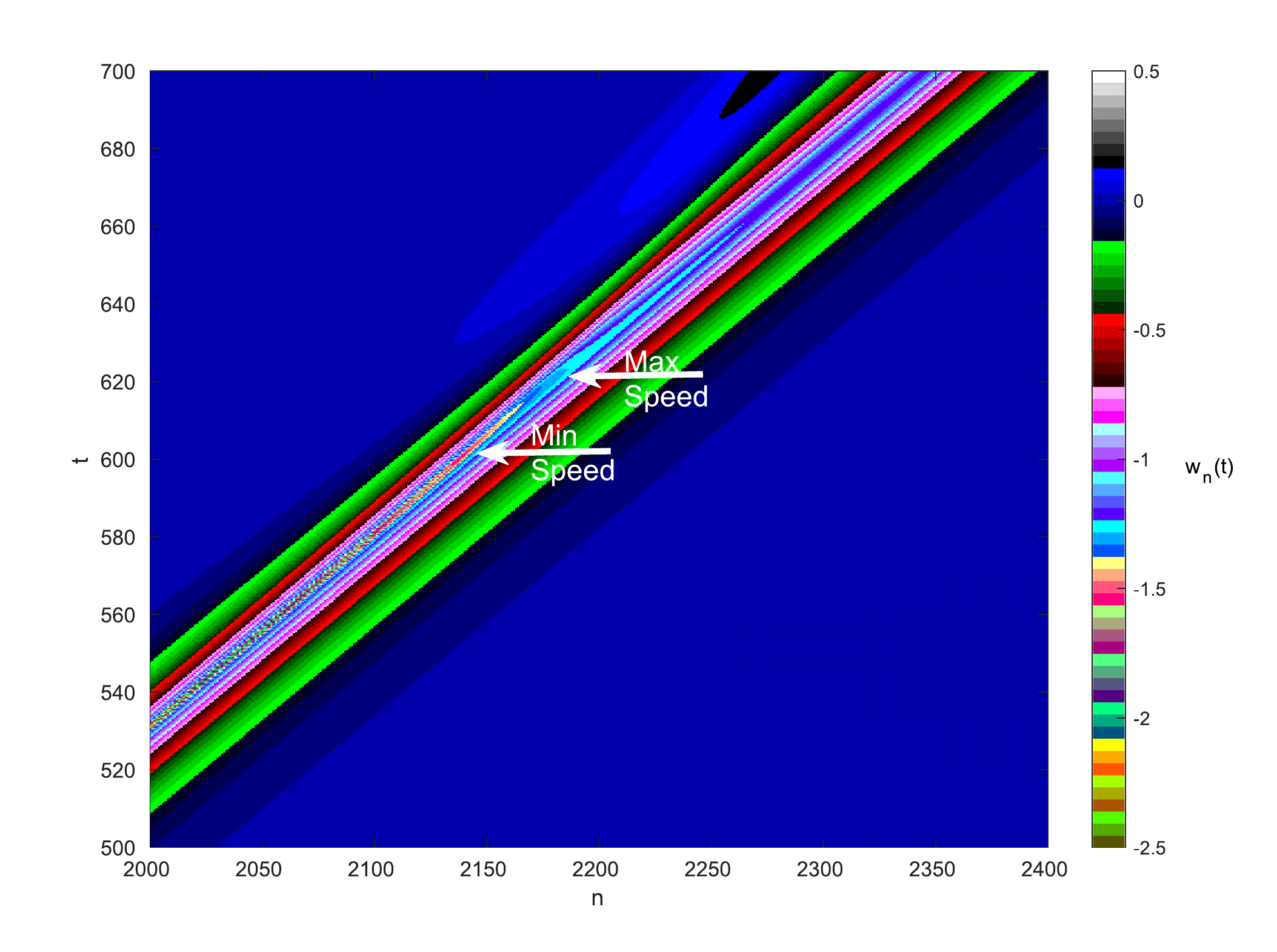}
\caption{\footnotesize An enlarged view of the space-time plot Fig.~\ref{fig:slow_evol_B}(a). The arrows mark the points corresponding to the minimal and maximal values of the wave's velocity in Fig.~\ref{fig:slow_cH_B}(a).}
\label{fig:slow_evol_B_zoom}
\end{figure}

The slowing-down case for the unstable STW with velocity $2.0785$ (point $B$ in Fig.~\ref{fig:Zregion}) is shown in Fig.~\ref{fig:slow_cH_B} and Fig.~\ref{fig:slow_evol_B}.
Note that the wave's velocity experiences a highly nonmonotone evolution in this case but eventually settles down to a lower value than the speed of the perturbed wave (point $B_1$ in Fig.~\ref{fig:Zregion}(a)), as can be seen in Fig.~\ref{fig:slow_cH_B}(a). Fig.~\ref{fig:slow_evol_B_zoom} zooms in the space-time plot of $w_n(t)$ in the time interval that includes times when the propagation velocity in Fig.~\ref{fig:slow_cH_B}(a) reaches its minimum and maximum. One can see that expulsion of the dispersive wave starts shortly after the velocity reaches its peak value.

When the sign of the perturbation is reversed, the wave speeds up after expelling a small-amplitude STW, and the ensuing dynamics is similar to the one shown in Fig.~\ref{fig:fast_cHp} and Fig.~\ref{fig:fast_evol} for the $N$-region. Similar slowing-down and speeding-up scenarios are observed for simulations perturbing the unstable wave that corresponds to point $A$ in Fig.~\ref{fig:Zregion}(a).

\section{Concluding remarks}
\label{sec:conclusions}

In the present work we have revisited the existence, stability and dynamical
features of lattice traveling waves in models where the competition between
short-range nonlinear interactions and longer-range linear ones may give
rise to stability changes. To this end, we considered the model where the nearest
neighbors feature an $\alpha$-FPU interaction, while interactions beyond nearest neighbors are
harmonic with exponentially decaying strength, and investigated
different parameter regimes. The regime where the strength and rate of decay of the longer-range interactions
were such that the energy $H$ of solitary traveling waves was a nonmonotone function of their velocity $c$
($N$-region) was observed to yield instability when $H'(c)<0$, in line
with earlier work. A more detailed study was also performed
in $Z$-region of the parameter space where $H(c)$ was not even single-valued. There, it was revealed that instability corresponds to $H'(s)<0$,
where $s$ is a parameter along the energy-velocity curve. In the Appendix we proved that the change in the sign of $H'(s)$ is sufficient for the change of stability.

A focal point of the present study concerned the dynamics of unstable solutions
in the regions where Floquet multipliers $\mu$ of the associated spectral stability
analysis were found to be $\mu > 1$. There, it was seen that it is possible to
``kick'' the unstable waveforms through suitable multiples of the
eigenvector associated with the instability to induce them to acquire a higher velocity, or recede to a lower speed.
In each of the cases, the velocity modification was accompanied by the concurrent
emission of a suitable coherent structure, typically represented by a slower pulse in the speeding-up case
and a dispersive shock wave when slowing-down. Such
possibilities were explored in both $N$ and $Z$ parameter regions.

Numerous questions arise as possible extensions of the present work
towards future study. In particular, it is important to understand
on a more general level what fundamental ingredients a physical setting must have
in order to induce the kind of competition that leads to $H'(c)<0$ and
the associated instability as is the case herein. An interesting and highly nontrivial extension of the present study in a
one-dimensional lattice setting would involve going beyond traveling
waves and examining breathers that bear a further internal frequency
(in addition to the traveling one). Finally, studies of solitary traveling waves in lattices
have been mostly limited to one-dimensional setting, and little is known about existence and stability
of such structures in higher dimensions. A systematic investigation of this issue in a suitably chosen model
would be a topic of interest in its own right.\\

\noindent {\bf Acknowledgements.} This work was supported by the U.S. National Science Foundation (DMS-1808956, AV and DMS-1809074, PGK) and by the
National Natural Science Foundation of China (NSFC-11801191, HX). We thank R. Pego for useful discussions and J. Cuevas-Maraver for sharing computer codes that were adapted to perform a number of computations presented herein.

\section*{Appendix: stability criterion}

In this Appendix, we generalize the stability criterion in \cite{FP02,FP04a,Cuevas17,Xu18} to the case when the energy of a STW is not necessarily a single-valued function of its velocity. As in \cite{Cuevas17,Xu18}, we consider a more general Hamiltonian than in \cite{FP02,FP04a} that goes beyond nearest-neighbor interactions. Although the analysis retraces the steps in these earlier works, with appropriate modifications, some issues still need to be addressed. In particular, in \cite{Cuevas17,Xu18} the effect of essential spectrum of the linearization operator was neglected in the derivation of the stability criterion and perturbation results, and the proof was provided for the displacement formulation, rather than the strain variables more appropriate for the problem at hand and also used in \cite{FP02,FP04a}. However, the choice of weighted spaces in \cite{FP02,FP04a} has resulted in the lack of skew symmetry of the associated symplectic form. In what follows, we use the strain formulation to provide a relatively complete argument for the generalized stability criterion by considering weighted spaces that eliminate the essential spectrum of the linearization operator and preserve the skew symmetry of the symplectic form.

\subsection*{Weighted spaces, skew symmetry and essential spectrum}
Consider a Hamiltonian system in the form
\beq
H=\sum_{n \in \mathbb{Z}}\left( \frac{1}{2}p_n^2+U(w_n)\right)=\sum_{n \in \mathbb{Z}}H_n(w(t),p(t))
\label{eq:Hamil2}
\eeq
where $U(w)$ is an interaction potential (that may include long-range interactions as in \eqref{eq:Ham}), $p(t)=[p_n(t)]$ is an infinite vector of particle momenta and $w(t)=[w_n(t)]$ is the strain vector. The dynamics of the lattice are governed by
\beq
\frac{d}{dt}r(t)=\mathcal{J}\frac{\partial H}{\partial r}, \quad
r(t)=\begin{pmatrix}
w(t) \\
p(t)
\end{pmatrix},
\quad
\mathcal{J}=\begin{pmatrix}
0 & e^{\partial}-I \\
I-e^{-\partial} & 0
\end{pmatrix}.
\label{eq:Dynam}
\eeq
Here $e^{\pm\partial}$ are the shift operators satisfying $\left(e^{\pm\partial} x\right)_i=x_{i \pm 1}$, and $I$ is the identity operator.
\begin{remark}
Clearly, $\mathcal{J}$ is invertible on space $\ell^2 \times \ell^2$, but the inverse is not bounded in this space because zero is in the essential spectrum of $\mathcal{J}$. We also note that an element in $\ell^2 \times \ell^2$ may have a preimage outside $\ell^2 \times \ell^2$ for operator $\mathcal{J}$. Using weighted spaces, one can make $\mathcal{J}$ a one-to-one function and change the essential spectrum of $\mathcal{J}$ so that its inverse is bounded. In particular, if $a>0$ and $\ell^2_a=\{ u: \sum_{j\in\mathbb{Z} } |u_j|^2 e^{2a j} <\infty \}$, the inverse of $\mathcal{J}$ on $\ell^2_a \times \ell^2_a$ is explicitly given by
\beq
\mathcal{J}^{-1}_{a,a}=\left(\begin{array}{cc}
   0 & -\sum_{k=1}^{\infty} e^{k\partial} \\
   -\sum_{k=0}^{\infty} e^{k\partial} & 0
\end{array} \right)
\label{eq:J_inv_1}
\eeq
We note that the inverse of $\mathcal{J}$  depends on the choice of the weighted space. For example, the inverses of $J$ on $\ell^2_{-a} \times \ell^2_{-a}$, $\ell^2_{a} \times \ell^2_{-a}$ and $\ell^2_{-a} \times \ell^2_{a}$ are
\beq
\begin{split}
\mathcal{J}^{-1}_{-a,-a}&=\left(\begin{array}{cc}
   0 & \sum_{k=0}^{-\infty} e^{k\partial} \\
   \sum_{k=-1}^{-\infty} e^{k\partial} & 0
\end{array} \right),
\quad
\mathcal{J}^{-1}_{a,-a}=\left(\begin{array}{cc}
   0 & \sum_{k=0}^{-\infty} e^{k\partial} \\
    -\sum_{k=0}^{\infty} e^{k\partial} & 0
\end{array} \right)\\
&\text{and} \quad
\mathcal{J}^{-1}_{-a,a}=\left(\begin{array}{cc}
   0 & -\sum_{k=1}^{\infty} e^{k\partial} \\
    \sum_{k=-1}^{-\infty} e^{k\partial} & 0
\end{array} \right),
\end{split}
\label{eq:J_inv234}
\eeq
respectively.
In particular, $\mathcal{J}^{-1}_{a,a}u=\mathcal{J}^{-1}_{-a,-a}u=\mathcal{J}^{-1}_{a,-a}u=\mathcal{J}^{-1}_{-a,a}u$ when $u\in (\ell^2_{a}\cap\ell^2_{-a})\times(\ell^2_{a}\cap\ell^2_{-a})$ and
$$
\left(\begin{array}{cc}
    \sum_{k=-\infty}^{\infty} e^{k\partial} & 0 \\
    0 & \sum_{k=-\infty}^{\infty} e^{k\partial}
\end{array} \right) u=0.
$$
\label{rmk:1}
\end{remark}
\begin{remark}
If one considers $\mathcal{J}$ on $\ell^2\times \ell^2$, then its adjoint is also viewed on $\ell^2\times \ell^2$, and in particular $\mathcal{J}^*=-\mathcal{J}$, which implies that $J$ is skew-symmetric. If $\mathcal{J}$ is defined on $\ell^2_a \times \ell_a^2$, then its adjoint $\mathcal{J}_{a,a}^*$ can be viewed as an operator defined on $\ell^2_{-a} \times \ell_{-a}^2$. Since we can also treat $\mathcal{J}$ as an operator $\mathcal{J}_{-a,-a}$ on $\ell^2_{-a} \times \ell_{-a}^2$, then
$$\langle u,\mathcal{J}_{a,a} v\rangle_{\ell^2\times\ell^2}+\langle \mathcal{J}_{-a,-a} u, v \rangle_{\ell^2\times\ell^2}=0$$
where $u\in\ell^2_{a} \times \ell_{a}^2$, $v\in\ell^2_{-a} \times \ell_{-a}^2$ and $\langle\cdot,\cdot\rangle_{\ell^2\times\ell^2}$ represents the inner product on $\ell^2 \times \ell^2$. This property can be equivalently written as $\mathcal{J}^*_{a,a}=-\mathcal{J}_{-a,-a}$. In fact, it can be shown that this type of skew-symmetric property holds for $\mathcal{J}$ on all of the four weighted spaces mentioned in Remark.\ref{rmk:1}.

However, since $\mathcal{J}^{-1}$ has different inverses on different weighted spaces, it in general does not inherit the skew symmetry from $\mathcal{J}$. For a weaker version, $\mathcal{J}^{-1}_{a,-a}$ and $\mathcal{J}^{-1}_{-a,a}$ satisfy
$$\langle u,\mathcal{J}^{-1}_{a,-a} v\rangle_{\ell^2\times\ell^2}+\langle \mathcal{J}^{-1}_{a,-a} u, v \rangle_{\ell^2\times\ell^2}=\langle u,\mathcal{J}^{-1}_{-a,a} v\rangle_{\ell^2\times\ell^2}+\langle\mathcal{J}^{-1}_{-a,a} u, v \rangle_{\ell^2\times\ell^2}=0$$
where $u,v\in (\ell^2_{a} \times \ell_{-a}^2)\cap(\ell^2_{-a} \times \ell_{a}^2)=(\ell^2_{a}\cap\ell^2_{-a})\times(\ell^2_{a}\cap\ell^2_{-a})$.
\label{rmk:2}
\end{remark}

We now assume that \eqref{eq:Dynam} has a smooth family of solitary traveling wave solutions which have the form
\beq
r_{tw}(t;s)=\begin{pmatrix}
w_{tw}(t;s) \\
p_{tw}(t;s)
\end{pmatrix}, \qquad
w_{tw,n}(t;s)=\hat{w}(\xi(s)), \qquad p_{tw,n}(t;s)=\hat{p}(\xi(s)),
\eeq
where $\xi(s) = n-c(s)t$ and $c(s)$ is the velocity of the wave, which is strictly above the sound speed and depends on the parameter $s$. We assume that $s$ provides a regular parametrization of the energy-velocity curve, so that $c'(s)$ and $H'(s)$ do not vanish simultaneously. This parametrization is not necessarily unique. It is convenient to use rescaled time $\tau = c(s)t$, so that the wave period is rescaled to one. Then we have
\beq
\frac{dR}{d\tau}=\frac{1}{c(s)}\mathcal{J}\frac{\partial H}{\partial R}, \quad R(\tau)=\begin{pmatrix}
   W(\tau) \\
   P(\tau)
\end{pmatrix}=r(t).
\label{eq:Dynam2}
\eeq
Linearizing \eqref{eq:Dynam2} around the solution $R_{tw}=\begin{pmatrix}
   W_{tw}\\
   P_{tw}
\end{pmatrix}$ with $R(\tau)=R_{tw}(\tau)+\epsilon S(\tau)$, we find
\beq
\frac{dS}{d\tau}=\frac{1}{c(s)}\mathcal{J}\frac{{\partial}^2 H}{\partial R^2}\bigg|_{R=R_{tw}}S(\tau).
\label{eq:Dynam3}
\eeq
We consider perturbations in the form $S(\tau)=S_{tw}(\tau)e^{\nu \tau}$, where
\[
S_{tw}=\begin{pmatrix}
   X_{tw} \\
   Y_{tw}
\end{pmatrix}
\]
is a traveling wave with unit velocity,
i.e. periodic modulo shift with period $1$. This yields the eigenvalue problem
\beq
\mathcal{L}S_{tw}(\tau)=\nu S_{tw}(\tau)
\label{eq:evalue}
\eeq
for the linear operator
\beq
\mathcal{L}:=\frac{1}{c(s)}\mathcal{J}\frac{{\partial}^2H}{\partial R^2}\bigg|_{R=R_{tw}}-\frac{d}{d\tau}
\label{eq:L_def}
\eeq
with eigenvalue $\nu$, which is related to the eigenvalue $\lambda$ used in the main text via $\nu=\lambda/c(s)$ due to the time rescaling. Note also that Floquet multiplier $\mu$ is related to $\nu$ via $\mu=e^\nu$. For the Hamiltonian \eqref{eq:Ham} the eigenvalue problem becomes
\beq
\begin{split}
-\frac{d}{d\tau}\left(\begin{array}{c}
   X_{tw,j}(\tau) \\
   Y_{tw,j}(\tau)
\end{array} \right)
+\frac{1}{c(s)}\left(\begin{array}{c}
   Y_{tw,j+1}(\tau)-Y_{tw,j}(\tau) \\
   V''(W_{tw,j}(\tau))X_{tw,j}(\tau)-V''(W_{tw,j-1}(\tau))X_{tw,j-1}(\tau)
\end{array} \right) \\
+\frac{1}{c(s)}\left(\begin{array}{c}
  0 \\
  \sum_{m=1}^{\infty}\Lambda(m) \Bigg{[} \sum_{l=0}^{m-1}X_{tw,j+l}(\tau) - \sum_{l=-m}^{-1}X_{tw,j+l}(\tau)  \Bigg{]}
\end{array} \right)
=\nu\left(\begin{array}{c}
   X_{tw,j}(\tau) \\
   Y_{tw,j}(\tau)
\end{array} \right).
\end{split}
\label{eq:evalue0}
\eeq
In order to investigate the case with well-localized perturbations $S_{tw}$ (that belong to spaces like $(\ell^2_{a}\cap\ell^2_{-a})\times(\ell^2_{a}\cap\ell^2_{-a})$) and exploit the skew-symmetric property of $\mathcal{J}^{-1}$, we view $\mathcal{L}$ as an operator densely defined on  $D_{tw,-a,a}^0([0,1])$ with domain $D_{tw,-a,a}^1([0,1])$, where
\begin{eqnarray*}
D_{tw,-a,a}^0([0,1]):=\Bigg{\{} Z(\tau)=\begin{pmatrix} X(\tau)\\ Y(\tau) \end{pmatrix},\tau \in [0,1] \Bigg| Z(1)= \left(\begin{array}{cc}
    e^{-\partial} & 0 \\
    0 & e^{-\partial}
\end{array} \right) Z(0),\\ \int_{0}^{1} \sum_{j \in \mathbb{Z}}  ({\lvert X_j(\tau) \rvert}^2 e^{-2a(j-\tau)}+{\lvert Y_j(\tau) \rvert}^2 e^{2a(j-\tau)}) d\tau < \infty \Bigg{\}}
\end{eqnarray*}
and
\begin{eqnarray*}
D_{tw,-a,a}^1([0,1]):=\Bigg{\{} Z(\tau)=\begin{pmatrix}   X(\tau)\\ Y(\tau) \end{pmatrix},\tau \in [0,1] \Bigg| Z(1)= \left(\begin{array}{cc}
    e^{-\partial} & 0 \\
    0 & e^{-\partial}
\end{array} \right) Z(0),\\
\int_{0}^{1} \sum_{j \in \mathbb{Z}}[({\lvert X_j(\tau) \rvert}^2+{\lvert X'_j(\tau) \rvert}^2) e^{-2a(j-\tau)}
+({\lvert Y_j(\tau) \rvert}^2+{\lvert Y'_j(\tau) \rvert}^2) e^{2a(j-\tau)}]d\tau < \infty \Bigg{\}},
\end{eqnarray*}
with prime denoting the time derivative.

Following the steps similar to the discussion about $\mathcal{J}$ on $\ell^2_{a}\times\ell^2_{-a}$,  we can also show that $\mathcal{J}$ has a bounded inverse on $D_{tw,-a,a}^0([0,1])$ (or $D_{tw,a,-a}^0([0,1])$) and $\mathcal{J}^{-1}_{-a,a}$ (or $\mathcal{J}^{-1}_{a,-a}$) is skew-symmetric due to
\[
\langle Z_g, \mathcal{J}^{-1}_{-a,a} Z_h\rangle+\langle\mathcal{J}^{-1}_{-a,a} Z_g, Z_h\rangle=0,
\]
with $\langle\cdot,\cdot\rangle$ being the inner product on $D^0_{tw,0,0}([0,1])$ and $Z_g, Z_h\in D_{tw,-a,a}^0([0,1])\cap D_{tw,a,-a}^0([0,1])$, where
\begin{eqnarray*}
D_{tw,a,-a}^0([0,1]):=\Bigg{\{} Z(\tau)=\begin{pmatrix} X(\tau)\\ Y(\tau) \end{pmatrix},\tau \in [0,1] \Bigg| Z(1)= \left(\begin{array}{cc}
    e^{-\partial} & 0 \\
    0 & e^{-\partial}
\end{array} \right) Z(0),\\ \int_{0}^{1} \sum_{j \in \mathbb{Z}}  ({\lvert X_j(\tau) \rvert}^2 e^{2a(j-\tau)}+{\lvert Y_j(\tau) \rvert}^2 e^{-2a(j-\tau)}) d\tau < \infty \Bigg{\}}
\end{eqnarray*}
and
\begin{eqnarray*}
D_{tw,a,-a}^1([0,1]):=\Bigg{\{} Z(\tau)=\begin{pmatrix}   X(\tau)\\ Y(\tau) \end{pmatrix},\tau \in [0,1] \Bigg| Z(1)= \left(\begin{array}{cc}
    e^{-\partial} & 0 \\
    0 & e^{-\partial}
\end{array} \right) Z(0),\\
\int_{0}^{1} \sum_{j \in \mathbb{Z}}[({\lvert X_j(\tau) \rvert}^2+{\lvert X'_j(\tau) \rvert}^2) e^{2a(j-\tau)}
+({\lvert Y_j(\tau) \rvert}^2+{\lvert Y'_j(\tau) \rvert}^2) e^{-2a(j-\tau)}]d\tau < \infty \Bigg{\}}.
\end{eqnarray*}

We note that when $\mathcal{L}$ is considered on unweighted spaces such as $D^0_{tw,0,0}([0,1])$, zero is usually embedded in the essential spectrum of $\mathcal{L}$.
To be specific, consider the Hamiltonian \eqref{eq:Ham}. Since $R_{tw}$ tends to zero and $V''(0)=1$, the limiting operator $\mathcal{L}_{\infty}$ can be defined as
\beq
\begin{split}
\mathcal{L}_{\infty}\left(\begin{array}{c}
   X_{tw,j}(\tau) \\
   Y_{tw,j}(\tau)
\end{array} \right)=
-\frac{d}{d\tau}\left(\begin{array}{c}
   X_{tw,j}(\tau) \\
   Y_{tw,j}(\tau)
\end{array} \right)
+\frac{1}{c(s)}\left(\begin{array}{c}
   Y_{tw,j+1}(\tau)-Y_{tw,j}(\tau) \\
   X_{tw,j}(\tau)-X_{tw,j-1}(\tau)
\end{array} \right) \\
+\frac{1}{c(s)}\left(\begin{array}{c}
  0 \\
  \sum_{m=1}^{\infty}\Lambda(m) \Bigg{[} \sum_{l=0}^{m-1}X_{tw,j+l}(\tau) - \sum_{l=-m}^{-1}X_{tw,j+l}(\tau)  \Bigg{]}
\end{array} \right)
\end{split}
\label{eq:L_inf}
\eeq
Substituting
\[
\left(\begin{array}{c}
X_{tw,j}(\tau)  \\
Y_{tw,j}(\tau)
\end{array} \right)=e^{ik(j-\tau)} \left(\begin{array}{c}
b_1   \\
b_2
\end{array} \right)
\]
into $\nu S_{tw}= {\mathcal{L}}_{\infty}S_{tw}$ and using $\Lambda(m)=J(e^{\alpha}-1) e^{-\alpha |m|}$, $m=1,2,\dots$, one can follow the procedure in \cite{FP04a} to compute the essential spectrum of $\mathcal{L}$ on $D^0_{tw,0,0}([0,1])$ in the form
\beq
\biggl\{\nu=i\biggl(k \pm \dfrac{2}{c(s)}\sin\dfrac{k}{2}\sqrt{1+\dfrac{J(e^{\alpha}+1)}{2(\cosh\alpha-\cos k)}}\biggr), \; k\in\mathbb{R} \biggr\}.
\eeq
Thus in this case the essential spectrum is along the imaginary axis and includes $0$. Similarly, the essential spectrum of $\mathcal{L}$ on $D^0_{tw,a,a}([0,1])$ with $a>0$ is obtained by replacing $k$ by $k+ia$ in the above, which yields
\beq
\begin{split}
&\biggl\{\nu=-a+ik \pm \dfrac{2i}{c(s)}\biggl(\cosh\dfrac{a}{2}\sin\dfrac{k}{2}+i\cos\dfrac{k}{2}\sinh\dfrac{a}{2}\biggr)\times\\
&\sqrt{1+\dfrac{J(e^{\alpha}+1)}{2(\cosh\alpha-\cos k\cosh a+i\sin k\sinh a)}}, \; k\in\mathbb{R} \biggr\}.
\end{split}
\eeq
One can show that for $c(s)>c_\text{s}$, where we recall that $c_\text{s}$ is the sound speed defined in \eqref{eq:sound}, the essential spectrum in this case is contained in the left half plane $\text{Re}(\nu)<0$ (and thus does not include zero) for $0<a<a_c$, where $a_c>0$ is the exponential decay rate of $R_{tw}$. It satisfies
\beq
\label{eq:ac}
\dfrac{2}{c(s)}\sqrt{1+\dfrac{J(1+e^\alpha)}{2(\cosh\alpha-\cosh a_c)}}\sinh\dfrac{a_c}{2}-a_c=0
\eeq
For $J \neq 0$, we have $0<a_c<\alpha$, with $a_c$ tending to zero as $c \to c_\text{s}$ and to $\alpha$ as $c \to \infty$. At $J=0$, $a_c$ solves
$2\sinh(a_c/2)=a_c c(s)$ \cite{FP04a}.

However, when the operator $\mathcal{L}$ is restricted to $D^0_{tw,-a,a}([0,1])$ (or $D^0_{tw,a,-a}([0,1])$) considered here, it has an empty essential spectrum because $\nu S_{tw}= {\mathcal{L}}_{\infty}S_{tw}$ has no solutions of form
\[
\left(\begin{array}{c}
X_{tw,j}(\tau)  \\
Y_{tw,j}(\tau)
\end{array} \right)=\left(\begin{array}{c}
e^{(ik+a)(j-\tau)} b_1   \\
e^{(ik-a)(j-\tau)} b_2
\end{array} \right).
\]

\subsection*{Zero eigenvalue and stability criterion}

We now differentiate \eqref{eq:Dynam2} with respect to $\tau$ to obtain
\beq
\frac{d^2 R}{d{\tau}^2}=\frac{1}{c(s)}\mathcal{J}\frac{{\partial}^2H}{\partial R^2}\frac{dR}{d\tau}.
\label{eq:EigenV1}
\eeq
Rearranging \eqref{eq:EigenV1} and evaluating it at $R=R_{tw}$ then yields $\mathcal{L}(\partial_{\tau}R_{tw})$=0. Thus $e_0:=\partial_{\tau}R_{tw}$ is an eigenvector of $\mathcal{L}$ with eigenvalue $\nu=0$ if $e_0\in D^0_{tw,-a,a}([0,1])$.  Multiplying \eqref{eq:Dynam2} by $c(s)$ and differentiating the result with respect to $s$, we obtain
\[
c'(s)\partial_{\tau}R+c(s)\partial_s\partial_{\tau}R=\mathcal{J}\frac{{\partial}^2H}{\partial {R}^2}\partial_s R
\]
Evaluating this equation at $R=R_{tw}$, we obtain
\beq
\mathcal{L}(c(s)\partial_s R_{tw})=c'(s)e_0,
\label{eq:e1}
\eeq
which for $c'(s) \neq 0$ yields
\[
\mathcal{L}(e_1)=e_0, \qquad e_1 := \frac{c(s)}{c'(s)}\partial_s R_{tw}.
\]
Thus $e_1$ is a generalized eigenvector of $\mathcal{L}$ for eigenvalue $\nu=0$ if $e_0, e_1\in D^0_{tw,-a,a}([0,1])$. Here we assume
\beq
e_0, e_1\in D_{tw,a,-a}^0([0,1])\cap D_{tw,-a,a}^0([0,1]),
\eeq
which holds when (positive) $a$ is less than $a_c$, the exponential decay rate of $R_{tw}$, which for our problem solves \eqref{eq:ac}.
This assumption then implies that the multiplicity of eigenvalue $\nu=0$ is always no less than two.
To further investigate the multiplicity of the eigenvalue $\nu=0$, we consider the adjoint of $\mathcal{L}$ as
\beq
\mathcal{L}^*=\frac{d}{d\tau}-\frac{1}{c(s)}\frac{\partial^2 H}{\partial R^2}\bigg|_{R=R_{tw}}\mathcal{J},
\eeq
on $D^0_{tw,a,-a}([0,1])$. Suppose that $\mathcal{L}_{a,-a}$ has the same form of $\mathcal{L}$, but it is restricted on $D^0_{tw,a,-a}([0,1])$. The adjoint of $\mathcal{L}$ for $Z \in D_{tw,a,-a}^0([0,1])$ can then be written as
\beq
\mathcal{L}^*Z=-\mathcal{J}^{-1}_{a,-a}\mathcal{L}_{a,-a}\mathcal{J}Z.
\eeq
Consider the generic case when $c'(s)\neq 0$ and $\text{ker}(\mathcal{L})=\text{span}\lbrace e_0 \rbrace$ and similarly $\text{ker}(\mathcal{L}^*)=\text{span} \lbrace \mathcal{J}^{-1}_{a,-a}e_0 \rbrace$, where we note that $\mathcal{L}^*(\mathcal{J}^{-1}_{a,-a} e_0)=0$. From the definition of $e_0$, it follows that
\beq
\mathcal{J}^{-1}_{a,-a} e_0=\mathcal{J}^{-1}_{-a,a} e_0.
\eeq
Using the skew-symmetry of $\mathcal{J}^{-1}_{a,-a}$ (see Remark~\ref{rmk:2}), one then can show that $\langle \mathcal{J}^{-1}_{a,-a} e_0, e_0\rangle=0$, and thus $e_0 \in (\text{ker}(\mathcal{L}^*))^{\perp}=\text{rng}(\mathcal{L})$ by the Fredholm alternative. Hence there exists $e_1$ such that $ \mathcal{L}(e_1)=e_0$.  Since the energy of the system is conserved, we have that
\[
H(s)=\int_{0}^{1} H|_{R_{tw}(\tau;s)}d\tau.
\]
We will use this to show that $H'(s)=0$ if and only if $\langle \mathcal{J}_{a,-a}^{-1}e_0, e_1\rangle=0$. Indeed,
\beq
\begin{split}
0 &= \left\langle e_1,\mathcal{J}_{a,-a}^{-1}e_0\right\rangle = \left\langle \frac{c(s)}{c'(s)}\partial_s R_{tw},\mathcal{J}^{-1}_{a,-a}\partial_{\tau}R_{tw}\right\rangle
= \left\langle c(s)\partial_s R_{tw},\frac{1}{c(s)}\frac{\partial H}{\partial R}\Bigg|_{R=R_{tw}}\right\rangle\frac{1}{c'(s)} \\
&=\frac{1}{c'(s)} \int_{0}^{1}\partial_s R_{tw}\left(\frac{\partial H}{\partial R}\right)\Bigg|_{R=R_{tw}}d\tau = \frac{1}{c'(s)}\int_{0}^{1} H'(s)|_{R=R_{tw}(\tau;s)}d\tau=\frac{H'(s)}{c'(s)}
\end{split}
\label{eq:Hprime}
\eeq
Thus, whenever $H'(s)=0$, we have that $e_1 \in \text{rng}(\mathcal{L})$, and hence there exists $e_2$ satisfying $\mathcal{L}(e_2)=e_1$, implying that the algebraic multiplicity of $\nu=0$ is at least three. Moreover, $\mathcal{L}^*(-\mathcal{J}^{-1}_{a,-a}e_1)=\mathcal{J}^{-1}_{a,-a}\mathcal{L}\mathcal{J} \mathcal{J}^{-1}_{a,-a}e_1=\mathcal{J}^{-1}_{a,-a}e_0$ implies that
\beq
\langle \mathcal{J}^{-1}_{a,-a}e_0, e_2 \rangle = \langle -\mathcal{L}^*\mathcal{J}^{-1}_{a,-a}e_1, e_2 \rangle =\langle \mathcal{J}^{-1}_{a,-a}e_1, \mathcal{L}e_2  \rangle=\langle \mathcal{J}^{-1}_{a,-a}e_1, e_1  \rangle=0,
\eeq
again by Remark~\ref{rmk:2}. Thus, if $s$ equals the critical value $s_0$ such that $H'(s_0)=0$ and $c'(s_0)\neq 0$,  there exists $e_3$ such that $\mathcal{L}e_3=e_2$, and the multiplicity of the eigenvalue $\nu=0$ is at least four.

\begin{remark}
The multiplicity of $\nu=0$ is closely related to the choice of space where $\mathcal{L}$ is defined. For example, for $\mathcal{L}$ defined on $D^0_{tw,a,-a}([0,1])$ or $D^0_{tw,-a,a}([0,1])$ the multiplicity of $\nu=0$ is at least four at $s=s_0$, while for $\mathcal{L}$ defined $D^0_{tw,a,a}([0,1])$ or $D^0_{tw,-a-a}([0,1])$ the multiplicity of the same eigenvalue at $s=s_0$ is just three because $\mathcal{J}^{-1}_{a,a} e_1\neq \mathcal{J}^{-1}_{-a,-a} e_1$ \cite{FP02,FP04a}.
\end{remark}

Since $\langle \mathcal{J}^{-1}_{a,-a}e_0, e_3 \rangle=\langle -\mathcal{J}^{-1}_{a,-a}e_1, e_2 \rangle$ does not generically vanish at $s=s_0$ where $H'(s_0)=0$, it is most common that $\nu=0$ has multiplictity four at $s=s_0$. The fact that $\nu=0$ has multiplicity four at $s=s_0$ and only two for $s\in (s_0-\delta, s_0+\delta)\backslash \{s_0\}$ is a sign for change of stability at $s=s_0$. In particular, if $H''(s_0)\neq 0$, we can use the perturbation procedure in \cite{Xu18} to obtain the following approximation of the near-zero eigenvalues of $\mathcal{L}$ near $s=s_0$:
\beq
\nu= \pm \sqrt{\frac{H''(s_0)(s_0-s)}{\langle \mathcal{J}^{-1}_{a,-a}e_0, e_3 \rangle} c'(s_0)}+O(|s-s_0|).
\label{eq:split}
\eeq
Thus, if $H''(s_0)c'(s_0)/\langle \mathcal{J}^{-1}_{a,-a}e_0, e_3 \rangle<0$, a pair of eigenvalues of order $O(|s-s_0|^{1/2})$ on the imaginary axis will collide at the origin and then split on the real axis as $s$ grows in $(s_0-\delta, s_0+\delta)$, which implies the emergence of instability. Similarly, if $H''(s_0)c'(s_0)/\langle \mathcal{J}^{-1}_{a,-a}e_0, e_3 \rangle >0$, the onset of instability will take place when $s$ crosses $s_0$ from above. This implies that if $c'(s_0)>0$ and $\langle \mathcal{J}^{-1}_{a,-a}e_0, e_3 \rangle>0$, an instability mode associated with a positive real eigenvalue emerges when the sign of $H'(s)$ changes from positive to negative at $s_0$.

\end{document}